\documentclass[final,3p,times,twocolumn]{elsarticle}

\usepackage{graphics}

\usepackage{amssymb}
\usepackage{amsmath}
\usepackage{textcomp}
\usepackage{url}
\usepackage{tabularx}

\biboptions{authoryear}

\journal{JQSRT}

\newcommand{\ee}[1]{$\cdot$10$^{#1}$}
\newcommand{\degree}{\ensuremath{^\circ}}

\begin{document}

\begin{frontmatter}



\title{IPRT polarized radiative transfer model intercomparison project -- phase A}
\author[lmu]{Claudia Emde}
\ead{claudia.emde@lmu.de}
\author[lei]{Vasileios Barlakas}
\author[lil]{C\'eline Cornet}
\author[col]{Frank Evans}
\author[ges]{Sergey Korkin}
\author[mri]{Yoshifumi Ota}
\author[lil]{Laurent~C.-Labonnote}
\author[gsf]{Alexei Lyapustin}
\author[tro]{Andreas Macke}
\author[lmu]{Bernhard Mayer}
\author[lei]{Manfred Wendisch}

\address[lmu]{Meteorological Institute, Ludwig-Maximilians-University,
  Theresienstr. 37, Munich, Germany}
\address[lei]{Leipzig Institute for Meteorology, University of Leipzig, Stephanstr. 3, Leipzig, Germany}
\address[lil]{Laboratoire d'Optique Atmosph\'erique, Universit\'e Lille, France}
\address[col]{University of Colorado, Boulder, CO 80309, USA}
\address[ges]{Universities Space Research Association, Columbia, Maryland, USA}
\address[mri]{Meteorological Research Institute, Tsukuba, Japan}
\address[gsf]{NASA Goddard Space Flight Center, Greenbelt, Maryland, USA}
\address[tro]{Leibniz Institute for Tropospheric Research, Permoserstr. 15, Leipzig, Germany}

\begin{abstract}
The polarization state of electromagnetic radiation scattered by
atmospheric particles such as aerosols, cloud droplets, or ice crystals
contains much more information about the optical and microphysical
properties than the total intensity alone. For this reason an
increasing number of polarimetric observations are performed from
space, from the ground and from aircraft.  Polarized radiative
transfer models are required to interpret and analyze these measurements
and to develop retrieval algorithms exploiting polarimetric observations. In
the last years a large number of new codes have been developed, mostly
for specific applications. Benchmark results are available for
specific cases, but not for more sophisticated scenarios including polarized surface
reflection and multi-layer atmospheres. The International Polarized
Radiative Transfer (IPRT) working group of the International Radiation
Commission (IRC) has initiated a model intercomparison project in
order to fill
this gap. This paper presents the results of the first phase A of the
IPRT project which includes ten test cases, from simple
setups with only one layer and Rayleigh scattering to rather
sophisticated setups with a cloud embedded in a standard
atmosphere above an ocean surface. All scenarios in the first phase A of
the intercomparison project are for a one-dimensional plane-parallel
model geometry. The commonly
established benchmark results are available at the IPRT website
(\url{http://www.meteo.physik.uni-muenchen.de/~iprt}). 
\end{abstract}


\end{frontmatter}


\section{Introduction}
\label{sec:intro}
An increasing number of remote sensing instruments measure the
polarization state of electromagnetic radiation. Therefore, 
polarized radiative transfer codes  
are required to interpret and analyze the measurements and to
develop retrieval algorithms. Sensors that measure polarization from
space are, e.g., the Polarization and Directionality of the Earth׳s
Reflectances (POLDER) instrument onboard PARASOL (Polarization and
Anisotropy of Reflectances for Atmospheric Sciences coupled with
Observations from a Lidar) \citep{deschamps1994} and the Thermal and
Near Infrared Sensor for
Carbon Observation Fourier-Transform Spectrometer (TANSO-FTS) on the Greenhouse gas
Observing SATellite GOSAT \citep{kuze2009}.
Future missions include
e.g. the Climate Absolute Radiance and Refractivity Observatory
(CLARREO) \citep{wielicki2013} and the Multi-Viewing Multi-Channel
Multi-Polarization Imaging mission (3MI) on METOP-SG (Meteorological
Operational Satellite - Second Generation).
All-sky imaging systems are available to measure 
the polarized radiance distribution; such systems are
described for instance by \citet{liu1997}, \citet{kreuter2009} 
and references therein. 
The Research Scanning Polarimeter (RSP) \citep{Cairns1999, Cairns2003} has been used
for ground-based as well as airborne aerosol measurements.
Other multi-channel polarimetric instruments are the Airborne
Multiangle SpectroPolarimetric Imager (AirMSPI) \citep{diner2012} and
the Observing System Including PolaRisation in the Solar Infrared
Spectrum (OSIRIS) \citep{auriol2008}.
The commercially available ground-based polarimeter, CE318-DP,
developed by CIMEL Electronic (Paris,
France) is now available at several AERONET stations \citep{Li2014}.

A large number of models for
polarized radiative transfer have been developed in the last years for
various specific applications. They mostly have been validated against
existing benchmark data; e.g., \citet{coulson1960} and \citet{nataraj2009} for
Rayleigh scattering; e.g., \citet{dehaan1987,WaubenW1994a,garcia1989} for
layers including aerosols; and \citet{Kokhanovsky2010} for cloud and
aerosol scattering including realistic phase matrices. 
More references to published benchmark results are given on the IPRT website,
section ``benchmark results''. However, all
existing benchmark results are limited to one or two plane-parallel
layers with an underlying Lambertian surface. To simulate the
measurements of the above mentioned sensors, far more realistic
settings are required. Reasonable height profiles of
molecules, aerosols and clouds should be taken into account. For
clear-sky atmospheres, a plane-parallel model geometry is a
reasonable approximation. When clouds are analyzed it is also
important to look into effects resulting from the geometrical structure of
clouds, commonly called 3D-effects, hence validated 3D vector
radiative transfer codes are required. In order to simulate limb
observations, fully spherical vector codes are needed. Polarization by
the surface must be considered, in particular for aerosol remote
sensing from space. 

In order to support model developers and to set standards for
polarized radiative transfer modeling the International Radiation Commission
(IRC) has established the working group ``International Polarized
Radiative Transfer'' (IPRT) which is charged by the task to provide benchmark data for 
polarized radiative transfer simulations for realistic atmospheric
setups as needed to simulate the current and future satellite,
airborne and ground-based polarimetric sensors. 
In order to establish this benchmark dataset a model intercomparison
project has been launched. This paper summarizes the results from the
first phase of the project. Six vector radiative transfer models from various
international institutions have participated. The models use different
approaches to solve the vector radiative transfer equation, among them
are deterministic approaches based on discrete ordinates or spherical
harmonics and also statistical approaches based on Monte Carlo
methods. The test cases in the first phase include simple one-layer
setups, cases with polarized surface reflection, and cases with
realistic height profiles of molecules and aerosol particles.

The focus of this intercomparison project is the establishment of
benchmark results, therefore all models were run in high accuracy
mode. For realistic applications with limited computational time 
the models are usually run with lower accuracy.  
The first intercomparisons between models showed several larger differences,
some of them due to model errors which have been fixed in the course of this
project. The participants were allowed to provide corrected or more
accurate data. Finally a very good agreement for all test cases has
been found for most models. The commonly
established benchmark results are available at the IPRT website
(\url{http://www.meteo.physik.uni-muenchen.de/~iprt}). 
The next phase of the intercomparison project will start soon with
focus on 3D radiative transfer.

\section{Radiative transfer models}

\begin{table*}[htbp]
  \centering
  \label{tab:rte_solvers}
  \caption{Overview of radiative transfer models}
  \smallskip
  \begin{tabularx}{1.0\hsize}{lp{3cm}lp{2.2cm}X}
    \hline
    model name & method & geometry & arbitrary \newline output altitude &
    references \\ \hline
    3DMCPOL & Monte Carlo & 1D/3D  & no & \citet{cornet2010,fauchez2014} \\
    IPOL & discrete ordinate & 1D  & no &
    \url{ftp://climate1.gsfc.nasa.gov/skorkin/IPOL/} \\
    MYSTIC & Monte Carlo & 1D/3D$^{(a)}$ & yes & \citet{mayer2009, emde2010} \\
    Pstar & discrete ordinate & 1D & yes & \citet{ota2010} \\
    SHDOM & spherical harmonics \newline discrete ordinate & 1D/3D  & yes & \citet{Evans1998} \\
    SPARTA & Monte Carlo & 1D/3D & no & \citet{Barlakas2014} \\ \hline
    \multicolumn{5}{l}{$^{(a)}$MYSTIC includes fully spherical
      geometry for 1D and 3D.} \\
  \end{tabularx}
\end{table*}

\subsection{3DMCPOL}
\label{sec:3DMCPOL}
3DMCPOL is a forward Monte-Carlo model for radiative transfer in
three-dimensional atmosphere. It can compute the reflected or transmitted
Stokes vector as well as upwelling and downwelling fluxes. Initially
developed for solar radiation \citep{cornet2010}, it was recently
extended to thermal radiation \citep{fauchez2014}. To save time and
for an accurate computation of radiances, it uses the Local Estimate
Method \citep{marshak2005,mayer2009}. The medium is
divided into voxels (3D pixels) with constant cloud and aerosol
optical properties, that are the extinction coefficient, the single
scattering albedo, the phase function and the cloud temperature. For
highly peaked phase functions, the truncation of \citet{potter1970}
is implemented and
we also added recently the variance reduction method of
\citet{buras2011}.
Atmospheric profiles including temperature, pressure and
absorption coefficient of a correlated k-distribution can also be
specified. The molecular scattering is computed automatically
according to the pressure profile. A depolarization factor can be
specified. To save substantial time, the absorption computation is
done following the Equivalence Theorem \citep{partain2000, emde2011}: 
the computation of radiative transfer trough the
scattering medium is done once and the radiances are attenuated
according to the absorption coefficient of the k-distribution along
the geometrical path of the photons. A heterogeneous surface can also
be specified with Lambertian reflection, ocean or snow bidirectional
function.  3DMCPOL applications concern mainly the cloud
heterogeneities effects on total and polarized radiances and the
errors on retrieved parameters from passive sensors. For example, for
the polarized and multi-angular radiometer POLDER3/PARASOL, 3DMCPOL
was used to study on synthetic data the bias for retrieved optical
thickness and effective radius \citep{cornet2013} and also to test
aerosol above cloud retrieval \citep{waquet2013}. Studies on the
thermal radiation were also conducted to assess the bias due to cirrus
heterogeneity on the brightness temperature measured by the radiometer
IIR/CALIPSO \citep{fauchez2014} and on the retrieved effective
optical thickness and effective diameters \citep{fauchez2015}.
For this intercomparison project 10$^8$ photons were used for
simulations without ocean and 10$^7$ photons for
simulations including an ocean surface. For ocean reflection, 
the polarized bidirectional reflectance contribution function 
and corresponding probability densities are computed at the beginning of the
simulation and during the calculation interpolations are done to
obtain the new direction and the contribution to the top of the
atmosphere. These interpolations increase computational time,
therefore less photons were used for the simulations with ocean.  

\subsection{IPOL}
\label{sec:IPOL}
IPOL is a radiative transfer code that computes Intensity and
POLarization of radiation reflected from or transmitted through the
Earth atmosphere over a reflecting surface. Radiation field inside the
atmosphere is not computed thus saving computation time and
memory. The code is suitable for remote sensing systems located on the
ground and aboard satellites or high altitude aircrafts. The code is
written in Fortran 90/95, requires external BLAS-LAPACK libraries, and
is freely available for downloading from
\url{ftp://climate1.gsfc.nasa.gov/skorkin/IPOL/}.

Following libRadtran and XRTM
(\url{http://reef.atmos.colostate.edu/~gregm/xrtm/}), IPOL will soon
incorporate several solvers for the vector radiative transfer
equation. This allows for fast yet accurate computation in a variety
of scenarios. In this intercomparision, only the discrete ordinates
solver is validated. The next solver to be included in the IPOL
project, SORD (Successive ORDers of scattering), currently undergoes
intensive testing. SORD is already available at
\url{ftp://climate1.gsfc.nasa.gov/skorkin/SORD/}.

In IPOL, the system of coordinates and direction of positive rotation
of the frame of reference is defined exactly following
\citet[p.~11 and Sec.~3.2]{hovenier2004}.
The Stokes vector is computed at arbitrary viewing
directions, except for the horizon, using the dummy-node technique
\citep{chalhoub2000}. Singular value decomposition is used to
solve the system of equations at Gauss and dummy nodes. Scaling
transformation \citep{karp1980} stabilizes the solution of the system
for an arbitrary atmospheric optical thickness. Layers with different
optical properties and a reflecting surface are bound together using
the matrix-operator method \citep{nakajima86, Plass1973, ota2010}.
Single scattering path radiance and reflection of
the direct solar beam from the surface are computed analytically. In
order to avoid errors in the aureole \citep{korkin2012}, none of
the phase function truncation techniques \citep{rozanov2010}
has been implemented so far. IPOL ignores atmospheric curvature, 3D
effects, and thermal emission.
We have tested IPOL against the vector codes APC \citep{korkin2013},
RT3 \citep{evans1991}, SCIATRAN \citep{rozanov2013}, and the published
results (see references in the Introduction). The radiative transfer
code SHARM \citep{lyapustin2005} was used to test the total
intensity. Scalar surface models and interface for IPOL were adapted
from SHARM as well.

IPOL was run with 16 streams (half-sphere) for all Rayleigh and the spherical
aerosol case, with 128 streams for the spheroidal aerosol cases,
and with 256 streams for all cloud cases to ensure high accuracy of benchmark results.
 
\subsection{MYSTIC}
\label{sec:MYSTIC}
The radiative transfer model MYSTIC (Monte-Carlo code for the
phYsically correct Tracing of photons in Cloudy atmospheres)
\citep{mayer2009} is a
versatile Monte-Carlo code for atmospheric radiative transfer which is
operated as one of several radiative transfer solvers of the
libRadtran software package \citep{mayer2005}.
The 1D version of MYSTIC is freely
available at \url{http://www.libradtran.org}. MYSTIC may be used to
calculate polarized solar and thermal radiances, and also for
irradiances, actinic fluxes and heating rates \citep{klinger2014}.
The model has been used
extensively to generate realistic synthetic measurements for the
validation of various retrieval algorithms for cloud and aerosol
properties \citep{davis2013, bugliaro2011}. Further application fields
are e.g. photochemistry \citep{suminska2012} or remote
sensing of exo-planets. MYSTIC allows the definition of arbitrarily
complex 3D clouds and aerosols, an inhomogeneous surface albedo and
topography. Polarized surface reflection is also included. The model
can be operated in fully spherical geometry \citep{emde2007}, 
hence it can also be used for limb sounding applications. 
Polarization has been included by combining various methods
\citep{emde2010}. The local
estimate method \citep{marchuk1980, marshak2005} has been adapted to
account for polarization, which is essential for accurate radiance
simulations. An importance sampling method similar to \citet{collins1972}
is used to sample the
photon direction after scattering or surface reflection, the
probability of which depends
not only on the scattering angle (as in scalar radiative transfer) but
also on the relative azimuth angle between incident and scattered
direction. Sophisticated variance reduction methods are included
\citep{buras2011} which allow to calculate unbiased radiances for scattering
media characterized by strongly peaked phase functions without any
approximations. It is also possible to calculate polarized radiances
in high spectral resolution efficiently \citep{emde2011}.
For all cloudless simulations shown in this intercomparison $10^8$ photons were run 
and for the simulations including clouds $10^7$ photons were
used. For clouds less photons were used because of the much larger
computational time due to multiple scattering. Even though only $10^7$
photons were used, the results are not too noisy because of the
sophisticated variance reduction methods included in MYSTIC. 

\subsection{Pstar}
\label{sec:Pstar}
The radiative transfer (RT) code, Pstar, has been developed to
simulate the polarized radiation field of a vertically inhomogeneous 1D
system as approximated by several homogeneous layers \citep{ota2010}.
Pstar has been used to simulate polarized solar and thermal
radiation as measured by satellite, and to develop an aerosol
retrieval algorithm, an atmospheric correction algorithm, and a vicarious
calibration system that include the polarization effect
(e.g. \citet{fukuda2013, murakami2013}). The RT scheme of Pstar is
constructed using the discrete ordinate method and the matrix operator
method. The discrete ordinate method is applied to each homogeneous
layer in order to obtain the reflection/transmission matrices and the
source vector of the layer. Then, the matrix operator method is
applied to all layers to obtain the radiation field of the
multi-layered system. The Stokes parameters at any interfaces between
the homogeneous layers as well as at the top of the atmosphere and at
any propagation direction are obtained by post-processing using the
source function integration technique. Finally, more accurate Stokes
parameters are obtained using the single scattering correction
procedure. This RT scheme is originally based on the formulations of
\citet{nakajima86, nakajima88}, which are implemented as the scalar
RT code series of System for Transfer of Atmospheric Radiation (STAR)
\citep{ruggaber1994}. \citet{ota2010} have extended the RT
formulation to express the polarized radiation field and 
implemented it in Pstar code. The extended RT scheme is constructed to be
flexible for a vertically inhomogeneous system including the oceanic
layers as well as the ocean surface. Accordingly, Pstar can be used to
simulate the radiation field in the coupled atmosphere-ocean system
including the polarization effect. Pstar computes all four Stokes
parameters in the vector mode, although only the total radiance (I) is
obtained in the scalar mode. Furthermore, the semi-vector mode that
computes the three Stokes parameters (I, Q, and U) on the basis of the
3x3 phase matrix approximation is available. The computation of eigen
solutions of the discrete ordinate method is one of the most time consuming
parts. In the vector mode, the direct decomposition method
\citep{ota2010}
is used in order to acquire the complex eigen solutions,
which is necessary to calculate the Stokes parameter V
accurately. However, in the scalar and semi-vector modes, a
square-root decomposition technique as described by \citet{nakajima86}
is invoked to obtain the real eigen solutions
efficiently. In the inter-comparison of this paper, the vector mode
was used to compute four Stokes parameters. In the Rayleigh scattering
cases, 15 streams were used for both single and multi-layer
conditions. In the aerosol and cloud scattering cases, 90 streams were
used for single-layer cases and 30 streams for multi-layer cases. 
The number of streams refer to the half-sphere. 

\subsection{SHDOM}
\label{sec:shdom}
The spherical harmonics discrete ordinate method (SHDOM) was developed
for unpolarized 3D atmospheric radiative transfer \citep{Evans1998}.
SHDOM is a non-Monte Carlo method in that the radiation field in the
domain is discretized and solved for iteratively.  The source function
is discretized with a spherical harmonics series for the angular aspect
and grid points in a cartesian geometry for the spatial aspect.  For
computational efficiency an adaptive grid is used in which addition grid
points may be added to the regular base grid where the source function
is changing rapidly, such as at illuminated cloud boundaries.  The
solution iterations consist of  1) transforming the source function from
spherical harmonics to discrete ordinates, 2) integrating the source
function along discrete ordinates to obtain the radiance field, 3)
transforming the radiance field to spherical harmonics, and 4) computing
the source function (including the scattering integral) efficiently in
spherical harmonics from the radiance.  A sequence acceleration method
is used to speed up convergence of the iterations.  For highly-peaked
phase functions the delta-M method is used and the output radiance is
computed with the TMS method of \citet{nakajima88}, in which the single
scattering contribution is calculated with the exact phase function
instead of the truncated spherical harmonics approximation.  SHDOM does
not implement higher order scattering corrections and thus does not
provide accurate results for highly-peaked (i.e. cloud) phase functions
in the solar aureole region.

The SHDOM model can calculate the radiance field from solar and/or
thermal emission sources of radiation.  The extinction and single
scattering albedo of the medium are specified on a 3D grid and
trilinearly interpolated between grid points.  Instead of specifying the
phase matrix at every grid point, to save memory, a table of expansion
coefficients for many phase matrices is input, and each grid point has a
specified index into the phase matrix table.  Once the SHDOM iterations
are completed, radiances in many directions on a grid at any height,
hemispheric fluxes, net fluxes, mean radiances, and net flux convergence
may be efficiently computed.  Several types of bidirectional reflection
distribution function (BRDF) models for the surface are implemented, and
their parameters may vary across the domain.  A k-distribution approach
is used to integrate across spectral bands. For large 3D domains SHDOM
may be run on multiple processors using the Message Passing Interface
\citep{pincus2009}. SHDOM is distributed from
{\tt http://coloradolinux.com/shdom/}.

Recently polarization capability was added to SHDOM using the real
generalized spherical harmonics method of \citet{doicu2013}.  Key pieces
of Adrian Doicu's VSHDOM research code were adapted for use in polarized
SHDOM.  The generalized spherical harmonics basis uses 4x4 matrices,
$\mathsf{Y}_{lm}(\mu,\phi)$ with 6 non-zero elements (including a 2x2
block for $Q$ and $U$). The angles $\mu=\cos(\theta)$ and $\phi$
describe the radiance direction, where $\theta$ is the
zenith angle and $\phi$ is the azimuth angle in polar coordinates. The elements of the
$\mathsf{Y}_{lm}$ matrix
are various Wigner d-functions in $\mu$ multiplied by Fourier functions
in $\phi$. The radiance and source function are represented by vectors
with $N_{stokes}=1$ (scalar), 3, or 4 elements, and thus the memory use
for a polarized calculation is about $N_{stokes}$ times that for a
scalar calculation. SHDOM has special purpose subroutines for the
unpolarized case ($N_{stokes}=1$) so the polarized code serves
efficiently for scalar calculations.  There are two polarized surface
reflection models: Fresnel surface with waves \citep{mishchenko1997} used
for ocean, and depolarizing modified-RPV (including the hotspot) with
polarizing Fresnel reflection from randomly oriented microfacets
\citep{diner2012} used for land.  When enough memory is available, the
surface BRDF is precomputed for all incoming and outgoing discrete
ordinates, greatly speeding up computation for uniform non-Lambertian
surfaces.  The polarized SHDOM distribution includes Mie and T-matrix
\citep{Mishchenko1998b} codes for generating SHDOM scattering tables from
spherical or spheroidal/cylindrical shaped particles, respectively.  The
six unique elements of the phase matrices are represented as series
expansions in Wigner d-functions (the $I-I$ and $V-V$ elements are
standard Legendre polynomials).

The SHDOM results shown below are for a high resolution run
with the number of discrete ordinates in zenith and azimuth angles of
$N_{mu}=128$ and $N_{phi}=256$ and the cell splitting accuracy of
0.00003. 

\subsection{SPARTA}
\label{sec:sparta}
The Solver for Polarized Atmospheric Radiative Transfer Applications
(SPARTA) is a new three-dimensional
(3D) vector radiative transfer model introduced in
\citet{Barlakas2014}. When finished it will become freely available. 
The model is based on the statistical Monte
Carlo method (in the forward scheme) and calculates column-response
pixel-based polarized radiances for 3D inhomogeneous
cloudless and cloudy atmospheres. Hence, it is well suited for use in
remote sensing applications. SPARTA is based on the established
scalar Monte Carlo model of the Institute for Marine Research at the
UNIversity of Kiel (MC-UNIK, \citealt{Macke1999}). MC-UNIK has been
extended to take into account the polarization state of the
electromagnetic radiation due to multiple scattering by randomly
oriented non-spherical particles, i.e., coarse mode dust particles or
ice particles. The SPARTA model considers a 3D Cartesian domain with a
cellular structure. The latter is divided into grid-boxes, characterized
by a volume extinction coefficient $\beta_{\mathrm{ext}}$
or a scattering coefficient $\beta_{\mathrm{sca}}$, a scattering phase
matrix {\bf P}($\Theta$) with a scattering angle $\Theta$, and a
single scattering albedo $\omega_{\mathrm{o}}$. Directions are
specified by the azimuth and zenith angles. 
Free path
lengths are simulated as outlined by \citet{marchuk1980} by random
number processes with attenuation described by the law of
Lambert. Scattering directions are calculated according to an
importance sampling method
\citep{collins1972,marchuk1980,emde2010}. Absorption is taken into
account by decreasing the initial photon weight by the integrated
absorption coefficient, along the photon path, according to Lambert's
law. The surface contribution is calculated assuming isotropic
reflection (Lambertian surface) or ocean reflection as outlined by
\citet{mishchenko1997}. In order to obtain precise radiance
calculations for each wavelength the so-called Local Estimate Method
(LEM) has been applied
\citep{collins1972,marchuk1980,marshak2005}.
Other variance reduction methods have not been implemented so far.

The selected number of
photons used for all the test cases in this intercomparison was
$10\textsuperscript{8}$.

\section{Definition of test cases}

\subsection{Model coordinate system and Stokes vector}

For all test cases the Stokes parameters, which are defined as time
averages of linear combinations of the electromagnetic field vector
\citep{chandrasekhar50, hansen1974, 
  mishchenko2002, wendisch2012}, are calculated:
\begin{eqnarray}
  \begin{pmatrix}
    I \\ Q \\ U \\ V
  \end{pmatrix} 
  = \frac{1}{2}\sqrt{\frac{\epsilon}{\mu_p}}
  \begin{pmatrix}
    E_l E_l^\ast + E_r E_r^\ast \\
    E_l E_l^\ast - E_r E_r^\ast \\
    -E_l E_r^\ast - E_r E_l^\ast \\
    i(E_l E_r^\ast- E_r E_l^\ast) 
  \end{pmatrix}
  \label{eq:stokes}
\end{eqnarray}
Here, $E_l$ and $E_r$ are the components of the electric field vector
paralle{\bf l} and perpendicula{\bf r} to
the reference plane respectively. The pre-factor on the right hand
side contains the electric permittivity $\epsilon$ and the magnetic
permeability $\mu_p$.

The model coordinate system is defined by the vertical (z-axis), the Southern
direction (x-axis) and the Eastern direction (y-axis). 
The Stokes vector is defined in the reference frame spanned by the
z-axis and the propagation direction of the radiation. 
The sign of Stokes parameters $U$ and $V$ depends on the definition of
the model coordinate system. The results shown in this paper are for
the coordinate system as defined in the books by
\citet{hovenier2004} and \citet{mishchenko2002}. The sign of $U$ and
$V$ changes when the viewing azimuthal angle definition is changed from
anti-clockwise to clockwise and also when the definition of the
viewing zenith angle is with respect to the downward normal instead of
the upward normal. 
The models IPOL, SHDOM and 3DMCPOL use the definition according to
\citet{hovenier2004}. SPARTA uses a
different coordinate system but the signs are consistent with
\citet{hovenier2004}. MYSTIC and Pstar also use different coordinate
systems and obtain opposite signs for $U$ and $V$, all results for
these Stokes components shown
in this paper have been multiplied by -1.
  
The position of the sun is defined by the vector pointing from the
surface to the sun position, i.e. the direction opposite to the
propagation direction of the incoming radiation.

The degree of polarization $P$ is defined as follows: 
\begin{equation}
  \label{eq:pol_deg}
  P=\frac{\sqrt{(Q^2+U^2+V^2)}}{I}
\end{equation}

The definition of the test cases can also be found at
\url{http://www.meteo.physik.uni-muenchen.de/~iprt}. 

\subsection{Test cases including a single layer}

The first set of test cases are for a single layer including 
different atmospheric constituents, i.e. molecules, aerosols and cloud
droplets. There are two cases including surface reflection, one is for
a Lambertian surface and the other includes an ocean reflectance
matrix. 

\subsubsection{A1 -- Rayleigh scattering}
\label{sec:a1_setup}

\begin{figure*}[t]
  \centering
  \includegraphics[width=.8\hsize]{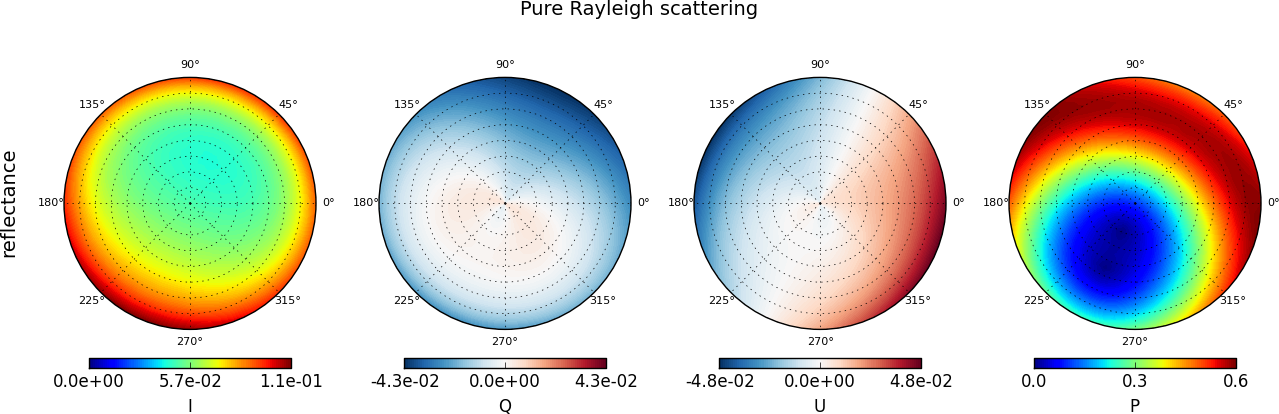}\\[1ex]
  \includegraphics[width=.8\hsize]{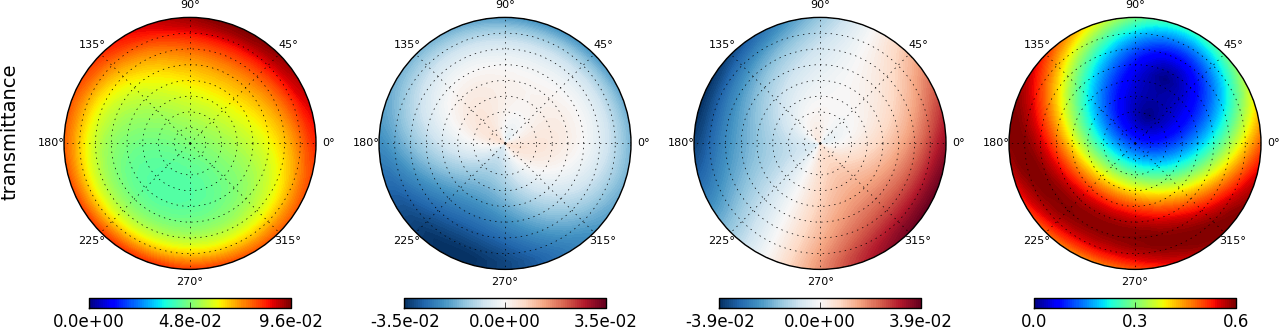}
  \caption{Results (MYSTIC)
    for Rayleigh scattering layer with an optical thickness of 0.5 and
    a depolarization factor of 0.1. The sun position is
    ($\theta_0$=30\degree, $\phi_0$=65\degree).
  }
  \label{fig:a1_rayleigh}
\end{figure*}

The most simple setup contains one layer with scattering
(non-absorbing) molecules. The radiation field is calculated at the
top and at the bottom of the layer for various sun positions and an
optical thickness of 0.5 (see Tab.~\ref{tab:a1_setup}). 
Viewing zenith angles range  from 0\degree\ to 80\degree\ at the bottom
and from 100\degree\ to 180\degree\ at the top of the layer with an increment of
5\degree. Viewing azimuth angles range from 0\degree\ to
360\degree\ with an increment of 5\degree.
This test is partly contained in
the tables by \citet{coulson1960, nataraj2009}. We also include
non-zero solar azimuth angles to test whether the models use consistent
coordinate systems to define the Stokes vector, this will be
particularly important for future intercomparisons in
three-dimensional geometry. Also we 
include a non-zero Rayleigh depolarization factor as defined in
\citep{hansen1974}, who defines the Rayleigh phase matrix as
follows: 
%
  \begin{align}
    \label{eq:rayleigh_phase_matrix}
    & {\bf P}(\Theta)  = \\
    & \begin{array}{l}
      \Delta \left[ 
        \begin{array}{cccc}
          \frac{3}{4}(1+\cos^2\Theta) & -\frac{3}{4}\sin^2\Theta &  0 & 0 \\
          -\frac{3}{4}\sin^2\Theta & \frac{3}{4}(1+\cos^2\Theta) &  0 & 0 \\
          0 & 0  & \frac{3}{2}\cos \Theta & 0 \\
          0 & 0 & 0 & \Delta' \frac{3}{2}\cos \Theta 
        \end{array}
      \right] \nonumber \\[6ex]
      \hspace{10ex}+(1-\Delta)\left[
        \begin{array}{cccc}
          \ 1\ &\ 0\ &\ 0\ &\ 0\ \\
          \ 0\ &\ 0\ &\ 0\ &\ 0\ \\
          \ 0\ &\ 0\ &\ 0\ &\ 0\ \\
          \ 0\ &\ 0\ &\ 0\ &\ 0\ 
        \end{array}
        \ \right],
    \end{array}
  \end{align}

where
\begin{eqnarray}
  \label{eq:depol}
  \Delta=\frac{1-\delta}{1+\delta/2}, \qquad 
  \Delta'=\frac{1-2\delta}{1-\delta},
\end{eqnarray}
and $\delta$ is the depolarization factor that accounts for the
anisotropy of the molecules. $\Theta$ is the scattering angle,
i.e. the angle between incoming and scattered directions. 
\begin{table}[h]
  \centering
  \begin{tabularx}{\columnwidth}{|X|X|X|}
    \hline
    solar zenith \newline angle $\theta_0$ & solar azimuth \newline angle $\phi_0$ &
    depolarization factor $\delta$
    \\ \hline
    0\degree & 65\degree & 0.0 \\
    30\degree & 0\degree & 0.03 \\
    30\degree & 65\degree & 0.1 \\
    \hline
  \end{tabularx}
  \caption{Sun positions and depolarization factors for test case A1,
    pure Rayleigh scattering layer.}
  \label{tab:a1_setup}
\end{table}
As an example Fig.~\ref{fig:a1_rayleigh} shows the radiation field for
$\theta_0$=30\degree, $\phi_0$=65\degree, and $\delta$=0.1. The $Q$-component of
the Stokes vector is negative in the principle plane. The
$U$-component is zero in the principal plane and it becomes positive
in the clockwise azimuthal direction and negative in the
counter-clockwise direction. The maximum degree of polarization
$P$ is clearly visible at a scattering angle of 90\degree.

\subsubsection{A2 -- Rayleigh scattering above Lambertian surface}
\label{sec:a2_setup}

This test case includes one layer with non-absorbing molecules with an
optical thickness of 0.1 above a Lambertian surface with albedo
0.3. The Rayleigh depolarization factor is 0.03 and the sun position
is $(\theta_0=50\degree, \phi_0=0\degree)$. The viewing directions are
as in Sec.~\ref{sec:a1_setup}. The only difference is that viewing
azimuths only range from 0\degree\ to 180\degree\ because the
radiation field is symmetric about the principal plane and the solar
azimuth angle is 0\degree. The top row in Fig.~\ref{fig:a_cases} shows the
transmittance (top part of polar plots) and the reflectance (bottom
part of polar plots). The transmitted radiation field is still highly polarized
whereas the degree of polarization $P$ becomes much smaller in the
reflected field (from $\sim$60\% in the maximum to $\sim$30\%) because the total intensity becomes
higher due to surface reflection. The order of magnitude of the Stokes
components $Q$ and $U$ is similar in transmitted and reflected
radiation fields. 

\subsubsection{A3 -- Spherical aerosol particles}
\label{sec:a3_setup}
\begin{figure}[h]
  \centering
  \includegraphics[width=1.\hsize]{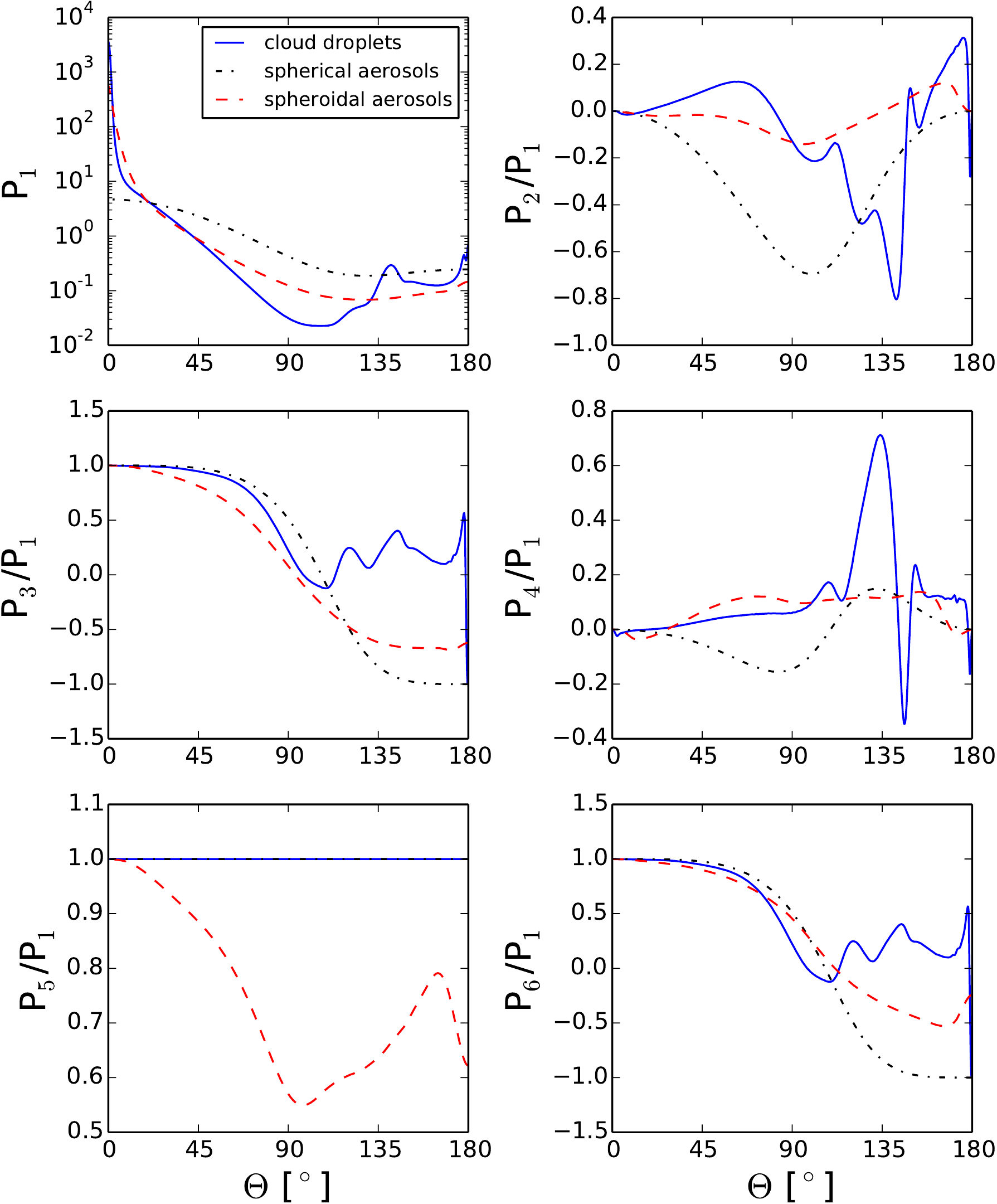}
  \caption{Phase matrices of spherical and spheroidal aerosol
    particles, and cloud droplets.}
  \label{fig:phamats}
\end{figure}
Here we calculate the transmitted and the reflected radiance fields
for a layer including spherical aerosol particles. The optical
thickness of the layer is 0.2 and the sun position is 
$(\theta_0=40\degree, \phi_0=0\degree)$.
The aerosol microphysical properties
correspond to typical water soluble aerosol for a relative humidity of
50\% at 350~nm as provided in the OPAC database \citep{hess1998,emde2010}: the complex
refractive index is 1.422-2.649\ee{-3}$i$, the size distribution is
log-normal with a mode radius of 26.2~nm and a width of 2.24, and the
mass density is 1.42~g/cm$^{-3}$. The optical properties including the
phase matrix $P$ are calculated using the Mie tool of the libRadtran
package \citep{mayer2005, wiscombe80a}. The phase matrix for
randomly oriented particles depends only
on the scattering angle $\Theta$:
\begin{eqnarray}
  \label{eq:a3_phase_matrix}
  \begin{array}{l}
    {\bf P}(\Theta) =  
    \left[ 
      \begin{array}{cccc}
        P_1(\Theta) & P_2(\Theta)  &  0 & 0 \\
        P_2(\Theta) & P_5(\Theta) &  0 & 0 \\
        0 & 0  & P_3(\Theta) & P_4(\Theta) \\
        0 & 0 & -P_4(\Theta) & P_6(\Theta)
      \end{array}
    \right] 
  \end{array}
\end{eqnarray}
The black dashed-dotted lines in Fig.~\ref{fig:phamats} show the phase matrix
elements $P_1$--$P_4$. For spherical particles it has only four
independent elements, $P_5$ is equal to $P_1$
and $P_6$ is equal to $P_3$. 
The expansion moments over generalized spherical functions (see \citet[Sec. 2.8]{hovenier2004})
for the phase matrix have also been made available for models which require
those as input. 

The radiance fields for this case are shown in the second row of
Fig.~\ref{fig:a_cases}. The maximum degree of polarization can be seen
at a scattering angle of approximately 100\degree. The $V$-component
of the Stokes vector is non-zero because scattering at spherical
droplets produces circular polarization. The size parameter of this
aerosol particles is small, therefore we do not see strong forward
scattering in the radiance field $I$.  

\subsubsection{A4 -- Spheroidal aerosol particles}
\label{sec:a4_setup}
A similar scenario as given in Sec.~\ref{sec:a3_setup} is calculated
for a layer including prolate spheroids with an aspect ratio of
3. Again a log-normal size distribution was assumed, with  a mode
radius of 390~nm and a width of 2. The complex refractive index is
1.52-0.01$i$ and the mass density is 2.6~g/cm$^{-3}$. 
The optical properties at 350~nm were calculated from a spheroid scattering data
base as described by \citet[Sec. 3.2]{gasteiger2011}. The scattering data base was
created using the T-matrix code by \citet{Mishchenko1998b} and the
geometric optics code by \citet{yang2007}.
When the asphericity of the aerosol particles is considered the
scattering phase matrix has six independent
elements. 
The red dashed lines in Fig.~\ref{fig:phamats} show the phase matrix elements for the
spheroidal particles. In contrast to the spherical aerosol particles
used in Sec.~\ref{sec:a3_setup} the phase function $P_1$ shows much
more forward scattering as the size parameter, i.e. the ratio between
particle size and wavelength, is larger. Also we see a
positive maximum in $P_2/P_1$ at a scattering angle of approximately
170\degree, whereas for Rayleigh scattering and for the small spherical
aerosol particles this ratio is always negative.  
For this case the sun position is $(\theta_0=40\degree,
\phi_0=0\degree)$ and the optical thickness is 0.2.

The third row in Fig.~\ref{fig:a_cases} shows the results for this
case. The reflected radiance field (upper half of the polar plots)
shows two maxima in the degree of polarization which are at the
scattering angles of the minima and maxima in the ratio $P_2/P_1$.
The transmitted radiance field shows high radiances $I$ in the
forward scattering region. The $Q$ and $U$ components show a
characteristic pattern in the forward scattering region.

\subsubsection{A5 -- Liquid water cloud}
\label{sec:a5_setup}
This test case includes a single layer with typical cloud droplets. The
scattering phase matrix (see blue lines in Fig.~\ref{fig:phamats}) has been
computed using the Mie tool of libRadtran for a wavelength of 800~nm 
assuming a gamma size distribution with an effective radius of 10~$\mu$m and a width of
0.1. The scattering phase matrix has the same structure as the one for
spherical aerosol particles (see Eq.~\ref{eq:a3_phase_matrix}). 
The cloud optical thickness is set to 5. This test case is used
to check whether features of the phase matrix, e.g. the cloudbow can be
simulated accurately. This is particularly important because
retrievals of the cloud effective radius from polarized observations
use the position and the width of the cloudbow \citep{breon2005,
  alexandrov2012}. Furthermore we want to check, whether the forward
scattering peak of highly asymmetric phase matrices can be taken into
account accurately by the models. The sun position is at
$(\theta_0=50\degree, \phi_0=0\degree)$.
The radiance is calculated at an
angular resolution of 1\degree\ in the principal plane
(i.e. $\phi\in[0\degree, 180\degree]$) at the top and at the bottom of
the layer. The same angular resolution is calculated in the almucantar
plane (i.e. $\theta=50\degree$ and $\phi\in[0\degree,1\degree,...,180\degree]$). 

\begin{figure}[h]
  \centering
  \includegraphics[width=1.\hsize]{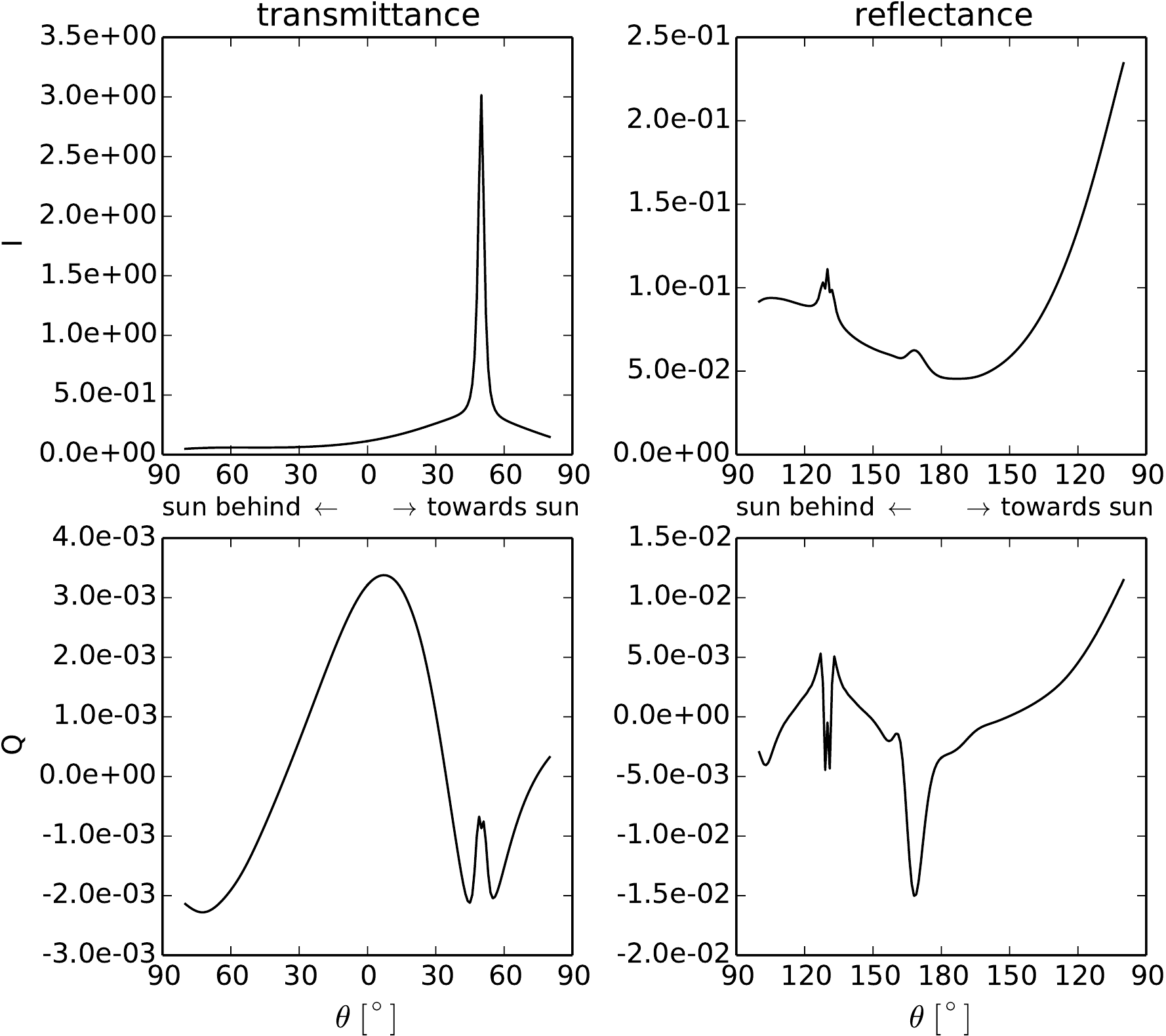}
  \caption{Transmitted and reflected radiance (IPOL simulations) for a
    cloud layer. The viewing directions $\theta$ are in the solar
    principal plane ($\phi\in[0\degree, 180\degree]$).}
  \label{fig:a5_principal}
\end{figure}

\begin{figure}[h]
  \centering
  \includegraphics[width=.9\hsize]{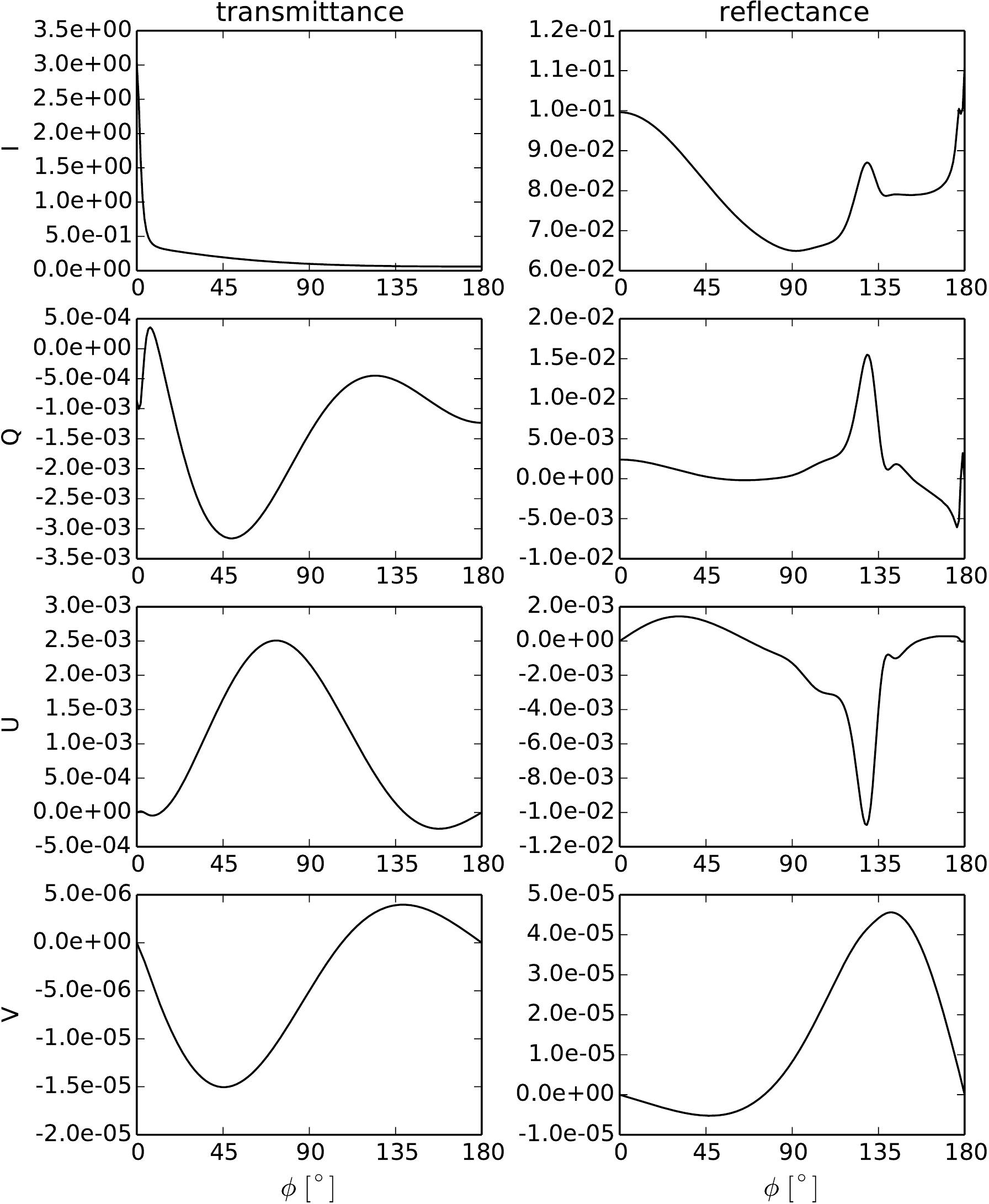}
  \caption{Transmitted and reflected radiance (IPOL simulations) for a
    cloud layer. Viewing directions are in the almucantar plane
    ($\theta=\theta_0, \phi\in[0\degree,1\degree,...,360\degree]$).}
  \label{fig:a5_almucantar}
\end{figure}

The reflected and transmitted radiances in the principal and
almucantar planes are shown in Fig.~\ref{fig:a5_principal} and
Fig.~\ref{fig:a5_almucantar} respectively.  The $I$-component of the Stokes vector
shows the strong forward scattering peak. The cloudbow can be seen in
the $I$- and much more pronounced in the $Q$-component of the Stokes
vector. The reason is that $Q$ is less affected by multiple scattering
than $I$, because photons that are scattered multiple times have
random polarization states.  

\subsubsection{A6 -- Rayleigh atmosphere above ocean surface}
\label{sec:a6_setup}
\begin{figure*}[t]
  \centering
  \includegraphics[width=1.\hsize]{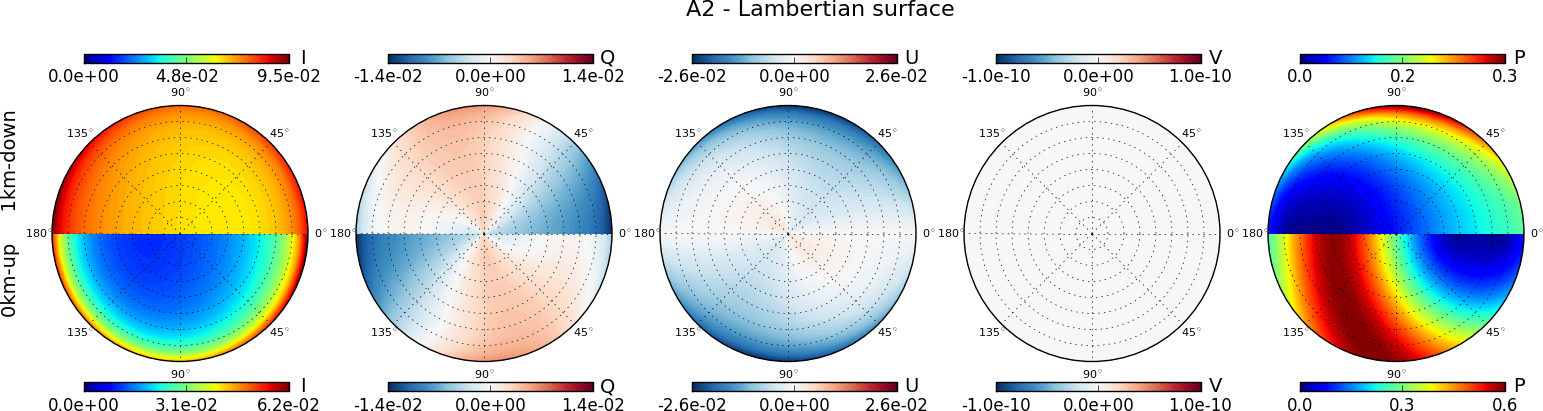}\\[0.8cm]
  \includegraphics[width=1.\hsize]{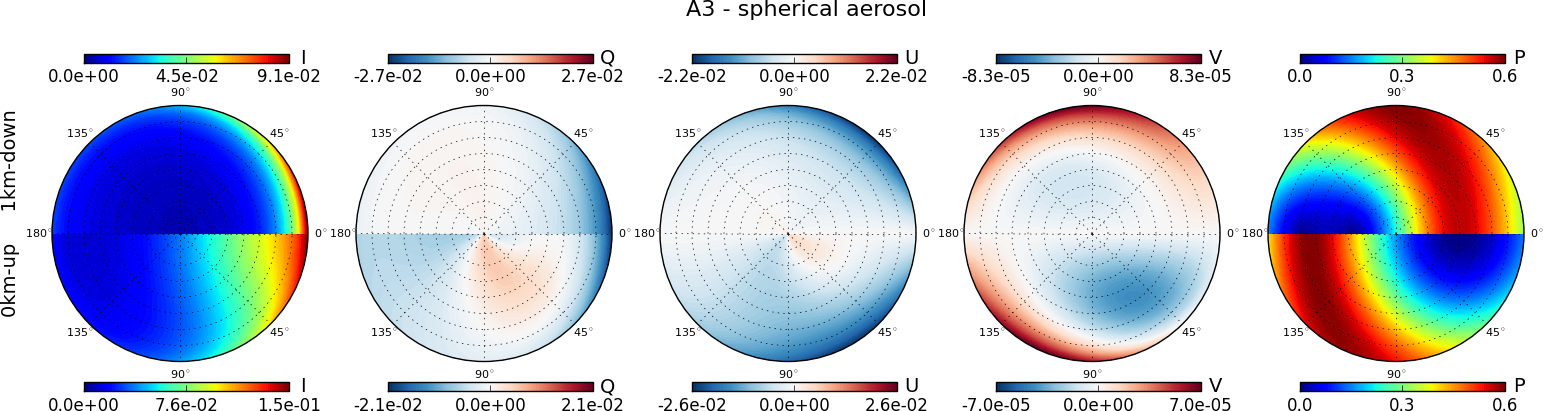}\\[0.8cm]
  \includegraphics[width=1.\hsize]{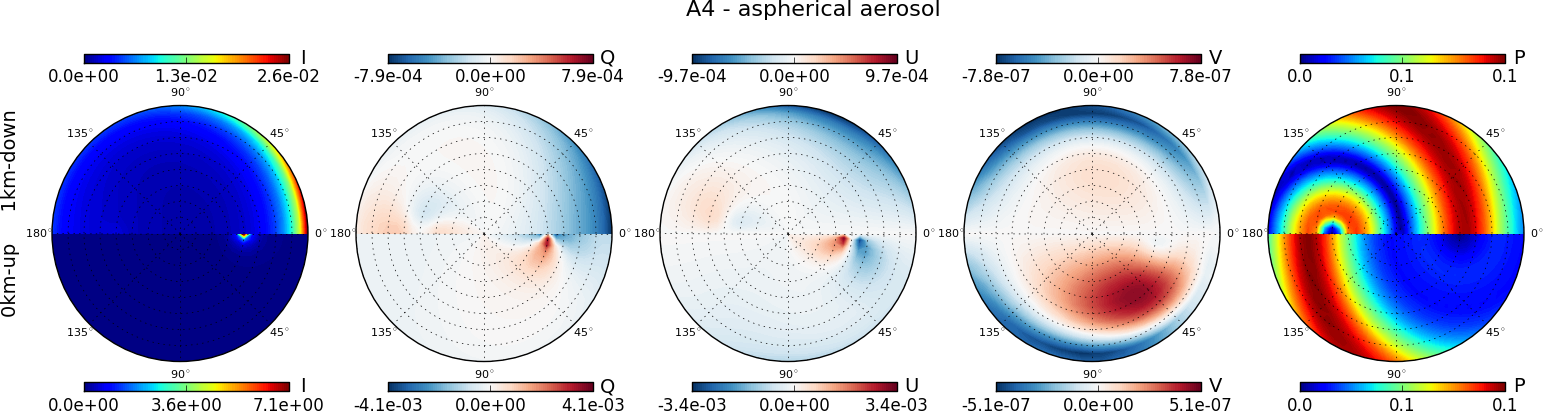}\\[0.8cm]
  \includegraphics[width=1.\hsize]{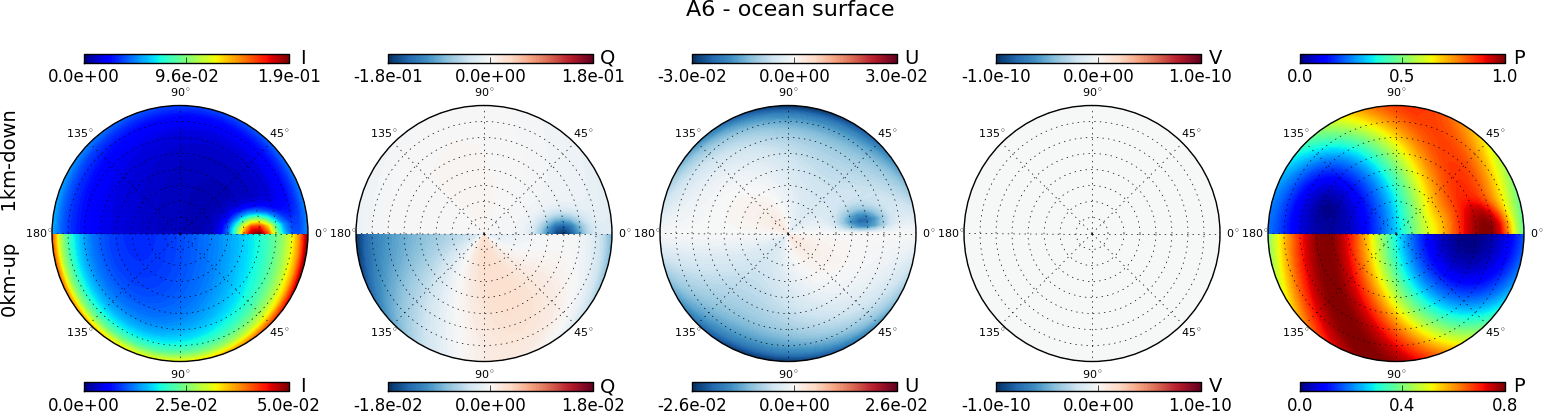}
  \caption{Results for Lambertian surface reflection (A2, 1st
    row), spherical aerosol particles (A3, 2nd row), aspherical aerosol
    particles (A4, third row), and ocean reflectance (A6, 4th row). For
    details about the setup please refer to the text,
    Secs.~\ref{sec:a2_setup}--\ref{sec:a6_setup}. The top part of the
    polar plots shows the Stokes components $I$, $Q$, $U$ and $V$ and the degree of
    polarization $P$ of the reflected radiance field $R$ and the bottom part
    shows the transmitted radiance field $T$. All results shown
    here are calculated using MYSTIC. 
    }
  \label{fig:a_cases}
\end{figure*}

In order to test whether the surface reflection matrix is correctly
included in the models, the radiance field is calculated for a
Rayleigh scattering layer (optical thickness 0.1, Rayleigh
depolarization factor 0.03) with an underlying ocean surface. The
reflectance matrix is calculated using a combination of Fresnel
equations and wave distribution including shadowing effects as
implemented by \citet{mishchenko1997}. The real part of refractive index of water
is assumed to be 1.33 and the imaginary part is zero. The wind speed
is assumed to be 2~m/s. The sun position is $(\theta_0=45\degree,
\phi_0=0\degree)$. 
The last row of Fig.~\ref{fig:a_cases} shows the results for this
case. The sun-glint is clearly visible in the reflected radiance (
$I$, $Q$- and $U$-components of the Stokes vector). Note that the
sign of $Q$ and $U$ of the reflection in the sun-glint is the same as
for Rayleigh scattering. 

\subsection{Test cases with realistic atmospheric profiles}

All following test cases are for the US-standard atmosphere
\citep{anderson1986} from 0 to 30 km altitude. The atmosphere is
divided into 30 layers with a thickness of 1~km. The radiance field is
calculated at the surface, at the top of the atmosphere and at an
altitude of 1~km. The radiance is calculated for viewing zenith angles
from 0\degree\ to 80\degree\ (up-looking) and from 100\degree\ to
180\degree\ (down-looking) and for viewing azimuth angles from 0\degree\
to 180\degree. The angular resolution in zenith and azimuth is
5\degree. The solar azimuth angle is generally 0\degree. The Rayleigh
depolarization factor is 0.03 and the surface albedo is 0 for all cases
apart from the case with ocean surface (Sec.~\ref{sec:b4_setup}).
\begin{figure}[h]
  \centering
  \includegraphics[width=1.\hsize]{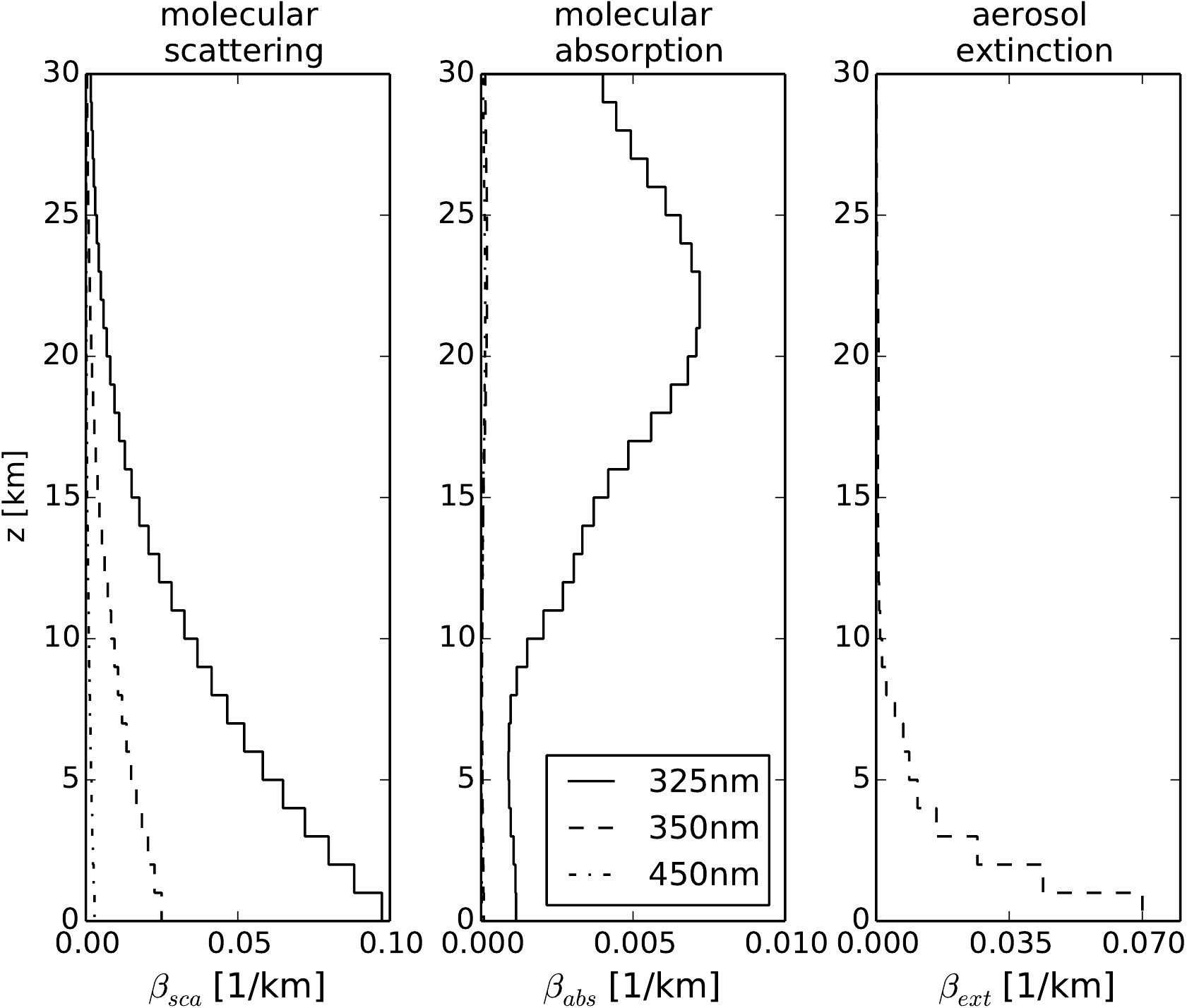}
  \caption{Scattering , absorption and extinction coefficient profiles for
    multilayer test cases. The left plot
    shows the molecular scattering coefficient, the middle plot
    the molecular absorption coefficient, and the right plot the
    aerosol extinction coefficient. Different lines correspond to
    case B1 (450~nm, dash-dotted), case B2 (325~nm, solid), and case B3
    (350~nm, dashed).}
  \label{fig:profiles}
\end{figure}

\subsubsection{B1 -- Rayleigh scattering for a standard atmosphere}
\label{sec:b1_setup}
The test case checks whether the discretization of the atmosphere into
plane-parallel layers is correctly implemented in the models. 
The radiance field is calculated at 450~nm taking into account
only Rayleigh scattering, i.e. molecular absorption is neglected. 
The sun position is $(\theta_0=60\degree, \phi_0=0\degree)$.
The scattering coefficient profile is shown in the left plot of
Fig.~\ref{fig:profiles}, the dash-dotted line corresponds to
450~nm. The Rayleigh depolarization factor is 0.03.
The upper two rows of Fig.~\ref{fig:B1_B2_results} show the radiance
field at top of atmosphere and surface and at 1~km altitude. As
expected the pattern is very similar to the simulation with one layer
(see Sec.~\ref{sec:a1_setup}). 

\subsubsection{B2 -- Rayleigh scattering and absorption for a standard atmosphere}
\label{sec:b2_setup}
\begin{figure*}[t!]
  \centering
  \includegraphics[width=1.\hsize]{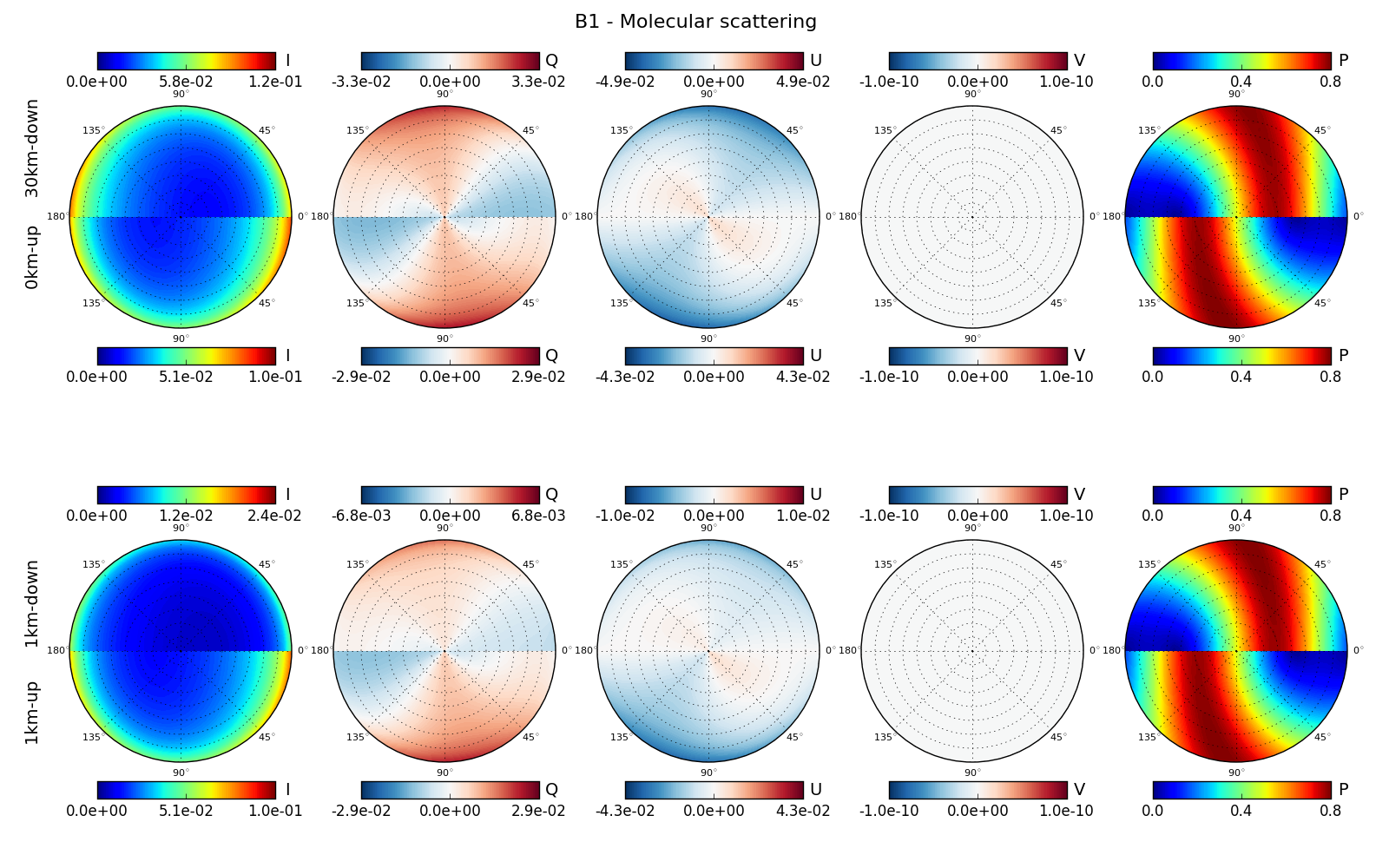}\\[0.3cm]
  \includegraphics[width=1.\hsize]{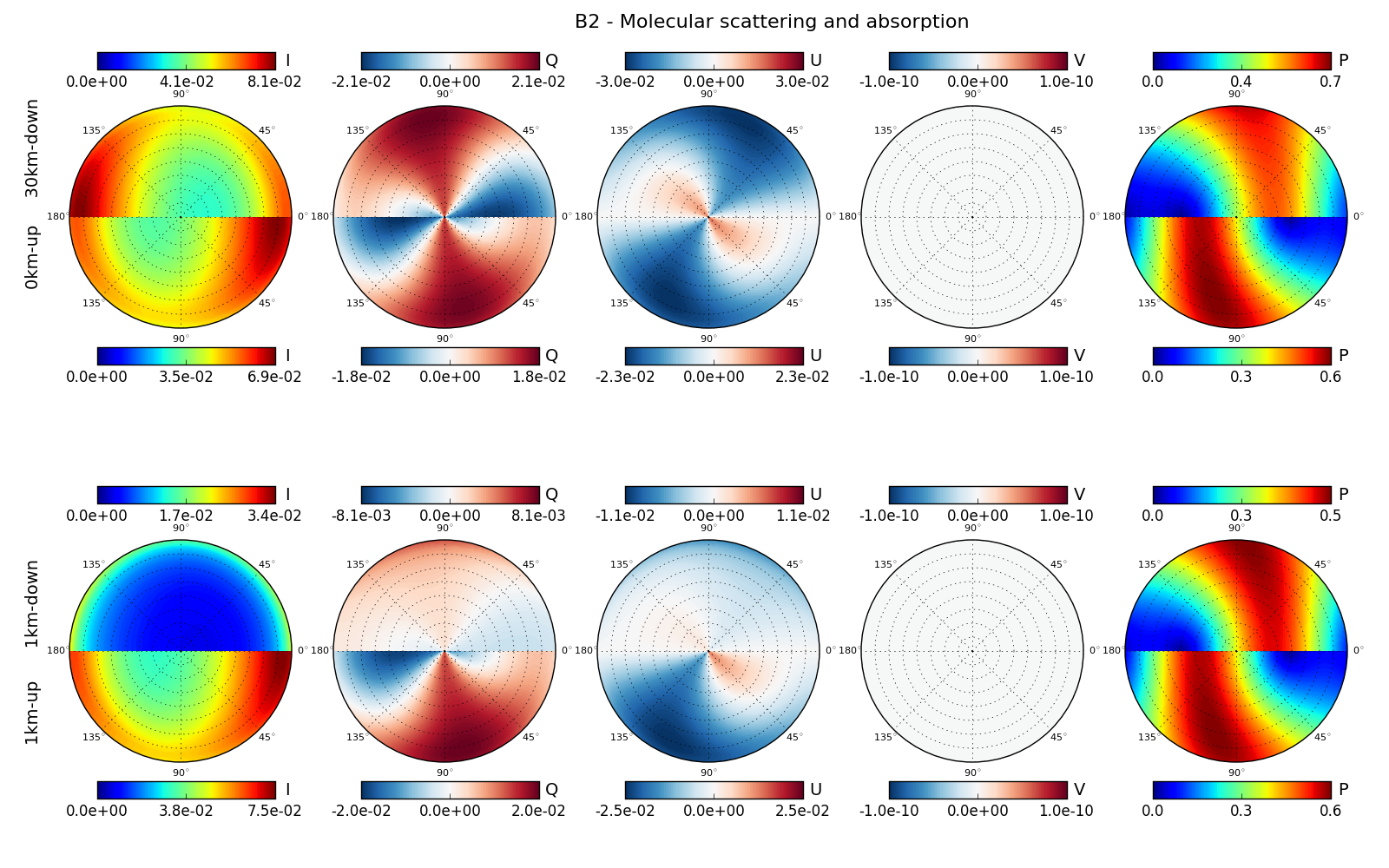}
  \caption{Radiation fields at surface (0km), top of atmosphere (30km),
    and at 1~km altitude. The upper two rows correspond to the pure
    Rayleigh scattering case (B1, see Sec.~\ref{sec:b1_setup}) and the lower 2
    rows are for Rayleigh scattering and absorption (B2, see
    Sec.~\ref{sec:b2_setup}).
    The upper parts of the polar plots are for
    downlooking directions and the lower parts for uplooking. The
    Stokes components and the degree of polarization are shown.} 
  \label{fig:B1_B2_results}
\end{figure*}

This case checks whether absorption is correctly taken into account. 
We calculate the radiance field at 325~nm for the US-standard
atmosphere. The sun position is $(\theta_0=60\degree,
\phi_0=0\degree)$. 
The scattering coefficient profile is shown in the left plot of
Fig.~\ref{fig:profiles} and the absorption coefficient profile
is shown in the middle plot; generally the solid lines corresponds to 325~nm.
Besides the strong absorption at this wavelength due to ozone Rayleigh scattering
is also much stronger than at 450~nm. 
The third and fourth row in 
Fig.~\ref{fig:B1_B2_results} show the radiance field at top of
atmosphere, at the  surface and at 1~km altitude. All Stokes components
show the characteristic Rayleigh scattering
pattern as for case B1 (pure Rayleigh scattering). However the degree
of polarization is smaller than for case B1. 

\subsubsection{B3 -- Aerosol profile and standard atmosphere}
\label{sec:b3_setup}
Here we check whether the models can correctly handle different
atmospheric constituents with similar extinction coefficients in the
same layers. 
We perform simulations at 350~nm for a standard
atmosphere including molecular absorption and scattering and
additionally an aerosol profile similar to \citet{shettle89} with a total optical
thickness of 0.2. Molecular absorption and scattering profiles and the
aerosol extinction profile are shown in Fig.~\ref{fig:profiles}
(dashed lines). We assume spheroidal aerosol partials with the same
optical properties as in Sec.~\ref{sec:a4_setup}. The sun position is $(\theta_0=30\degree,
\phi_0=0\degree)$. The upper two rows of Fig.~\ref{fig:B3_B4_results}
show the radiance field at top of atmosphere, at the surface and at 1~km
altitude. The total intensity $I$ for up-looking directions is dominated
by the forward scattering peak. The polarization pattern is dominated by
Rayleigh scattering and the features in the
radiation field for test case A4 (layer with aspherical aerosol
particles, see Fig.~\ref{fig:a_cases}), e.g. the two maxima in the
degree of polarization or the patterns in $U$ and $V$ in the forward
scattering region, are no longer visible. The main effect of aerosol
is a decrease in the degree of polarization which is most likely due to
the dilution of the strong Rayleigh polarization by the weaker aerosol
polarization.
 
\subsubsection{B4 -- Cloud above ocean surface}
\label{sec:b4_setup}
\begin{figure*}[t!]
  \centering
  \includegraphics[width=1.\hsize]{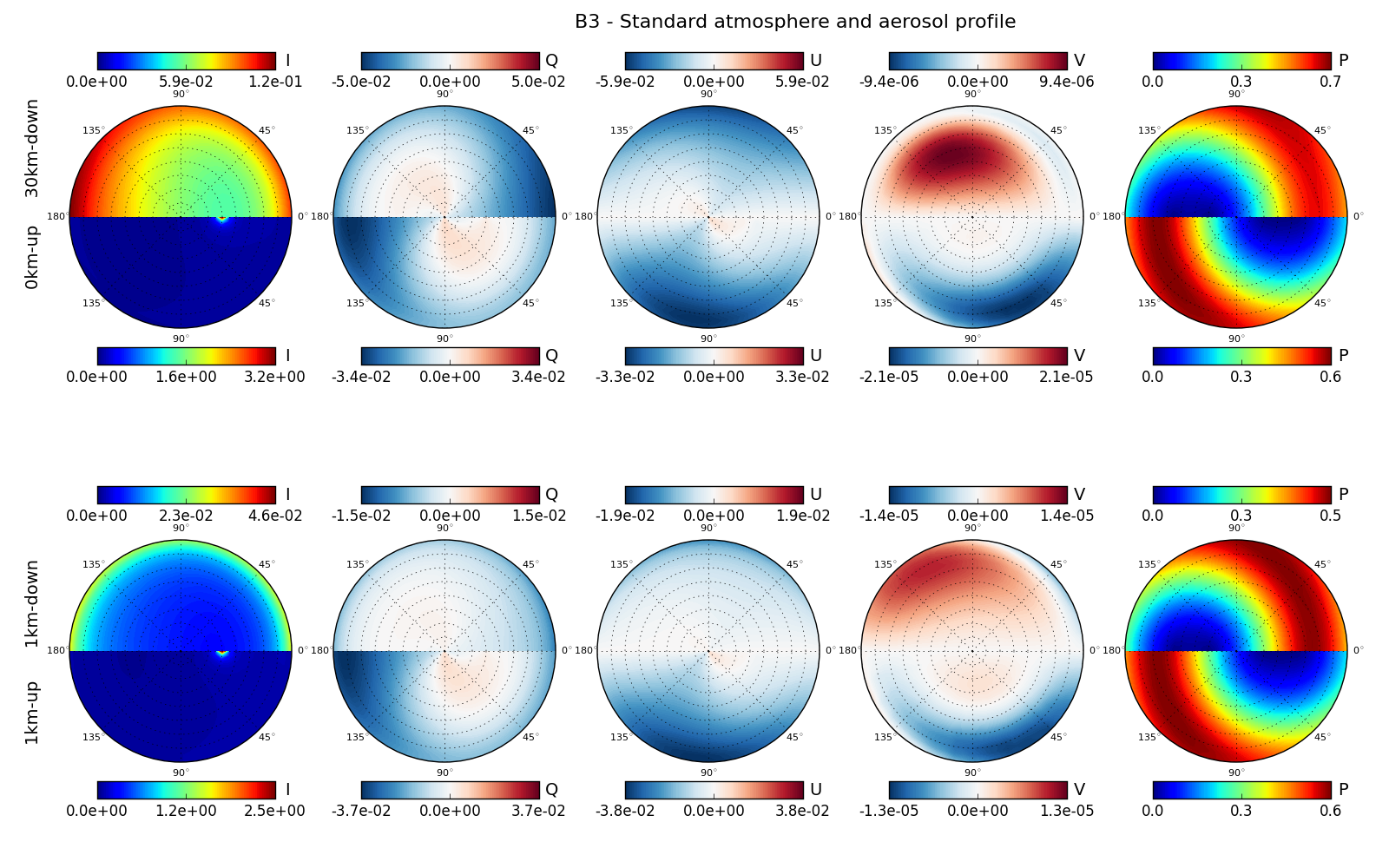}\\[0.3cm]
  \includegraphics[width=1.\hsize]{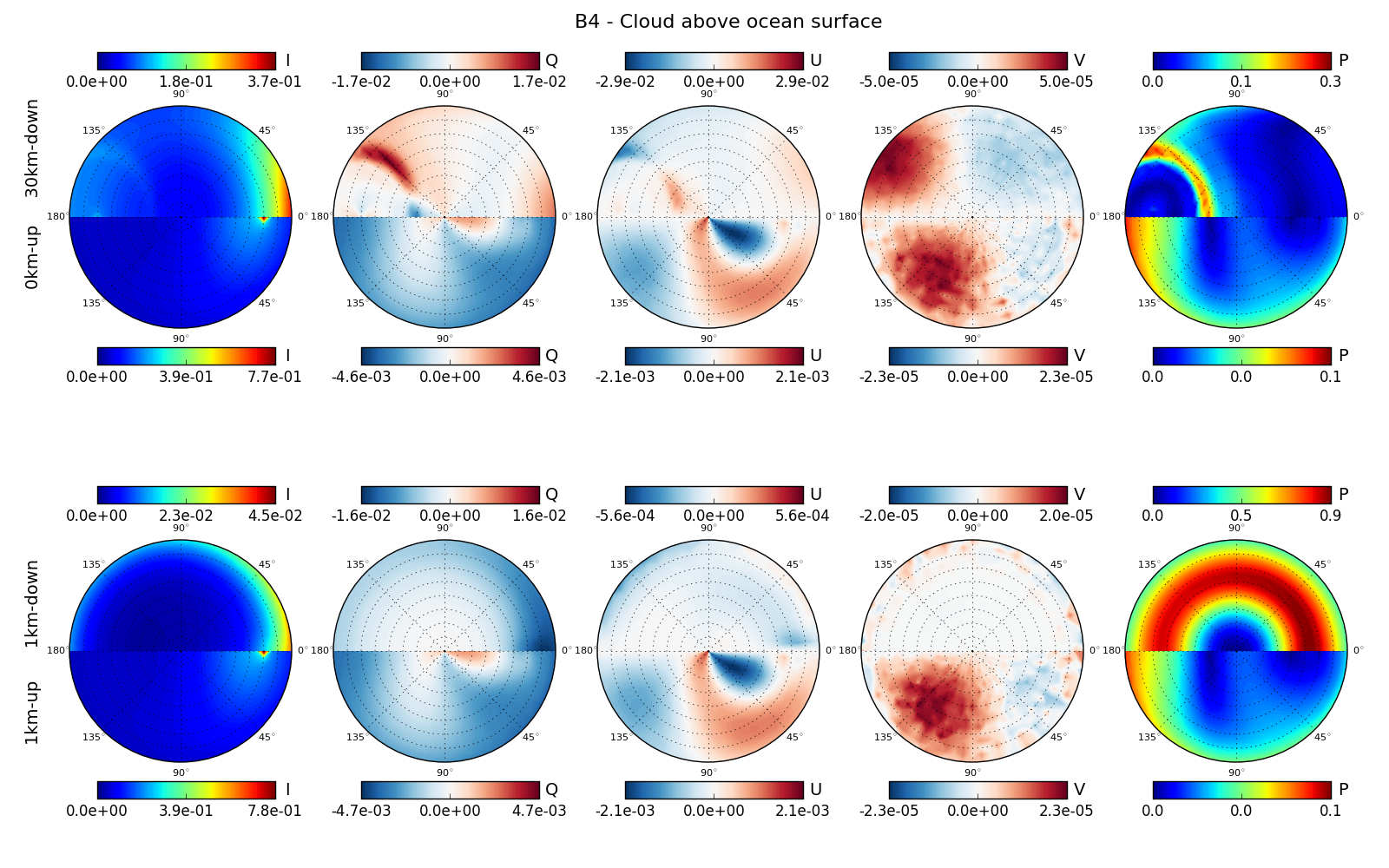}
  \caption{Radiation fields at surface (0km), top of atmosphere (30km),
    and at 1~km altitude. The upper two rows correspond to the case
    with aerosol profile (B3, see Sec.~\ref{sec:b3_setup}) and the lower 2
    rows are for a cloud layer above an ocean surface (B4, see
    Sec.~\ref{sec:b4_setup}).
    The upper parts of the polar plots are for
    downlooking directions and the lower parts for uplooking. The
    Stokes components and the degree of polarization are shown.} 
  \label{fig:B3_B4_results}
\end{figure*}

This most sophisticated test case includes a cloud layer embedded in a
standard atmosphere above an ocean surface.
The calculation is performed at a wavelength of
800~nm, molecular scattering is included, absorption is neglected. 
The ocean surface is defined as in
Sec.~\ref{sec:a6_setup}, here also we assume a wind speed of 2~m/s
and we use 1.33+0$i$ as the refractive index for water. 
Additionally, a cloud layer with an optical thickness of 5 is included
from 2~km to 3~km altitude. The cloud optical properties are the same
as in Sec.~\ref{sec:a5_setup}. The sun position for this case is  $(\theta_0=60\degree,
\phi_0=0\degree)$.
The lower two rows of Fig.~\ref{fig:B3_B4_results}
show the radiance field at top of atmosphere, at the surface and at 1~km
altitude. The down-looking radiance field (reflectance) at the top of
the atmosphere shows
the cloudbow very clearly with high contrast in the degree of
polarization, whereas in the total intensity the feature is very
weak. The degree of polarization for down-looking directions at 1¨km
altitude shows an interesting feature, it is $\sim$90\% for
a viewing angle of $\sim$53\degree\ corresponding to the
Brewster angle for the water surface ($\theta_B=\arctan(1.33)$). Due to multiple
scattering in the cloud layer we get incident radiation on the
water surface from all directions. Now the radiation which hits the surface
at the Brewster angle is fully polarized after reflection, and the
reflected direction is at the same angle, this nicely
explains the observed pattern. The ``ring'' is smeared because the
ocean surface with waves is not an ideal mirror.  
As for the aerosol case the total intensity for the up-looking
directions at the surface and at 1~km altitude 
is dominated by sharp forward scattering peak. The patterns for $Q$
and $U$ for up-looking directions show a characteristic cloud
scattering pattern which is very different from the Rayleigh
scattering pattern.

\section{Model intercomparison}
\label{sec:model_intercomparison}
This section presents results of all models for an exemplary selection
of viewing angles. For each test case, one or two plots show the
simulated Stokes vector and the absolute differences between MYSTIC
and IPOL, SPARTA, Pstar, SHDOM, and 3DMCPOL respectively. 
Out of range values are shown as arrows in the difference plots. 
Comparison plots for all viewing directions are provided on the IPRT
website and as supplementary material.
  
In order to quantify the level of agreement between the models we
calculate the relative root mean square differences
$\Delta_m$ between MYSTIC and other codes for the full radiation
field including all up- and down-looking directions. This yields one
representative number for each test case.
We define $\Delta_m$ for $I$ as follows:
\begin{equation}
  \label{eq:rel_diff}
\Delta_m=\frac{\sqrt{\sum_{i=1}^{N}{\left(I^{i}_{\rm MYSTIC}
- I^{i}_ {m}\right)^2}}}
{\sqrt{\sum_{i=1}^{N}{\left(I^{i}_{\rm MYSTIC}\right)^2}}}
\end{equation} 
Here $m$ denotes the radiative transfer model and the summation is
done over all $N$ directions, for which the radiation field is
calculated. For the other Stokes components $\Delta_m$ is calculated
accordingly.
We look at relative root mean square differences because the
Stokes components $Q$, $U$ and $V$ are differences between intensities
(see Eq. \ref{eq:stokes}) and for some geometries they have zero or
extremely small values. Mean relative differences are therefore not
meaningful because relative differences
for radiance values very close to zero become very large. 
A weakness of the definition of $\Delta_m$ is that it
might be dominated by a few very large differences at specific viewing
directions. 
For cases with strongly peaked phase functions (i.e. cloud cases)
$\Delta_m$ is dominated by the forward scattering region. Therefore we
also calculated $\Delta_m$ without the solar aureole region, 
i.e. directions up to 10\degree\ from the sun direction are taken out
of the summation in Eq.~\ref{eq:rel_diff}. 

\subsection{Test cases including a single layer}

The relative root mean square differences for the single layer test cases are
listed in Table~\ref{tab:a_results}, mostly the level of agreement is
of the order of 0.1\%. For the cloud case A5, the MYSTIC reference results,
especially for circular polarization, are more noisy, hence the
relative root mean square difference becomes larger although most
models still agree perfectly within the expected accuracy range. 

\begin{table*}[t!]
  \centering
  \begin{tabular}{|l l|c|c|c|c|c|c|c|c|c|}
    \hline
model name &  & A1 & A2 & A3 & A4 & A5$^{\rm pp}$ & A5$^{\rm pp}_{\rm part}$ & A5$^{\rm al}$ & A5$^{\rm al}_{\rm part}$ & A6 \\ \hline \hline 

      IPOL & I &    0.017 &    0.009 &    0.102 &    0.088  &    0.183 &    0.077 &    0.178 &    0.075&    0.064 \\ 
           & Q &    0.024 &    0.036 &    0.287 &    0.028  &    0.794 &    0.771 &    0.816 &    0.803&    0.124 \\ 
           & U &    0.029 &    0.029 &    0.295 &    0.034  &       -  &       -  &    1.058 &    1.039&    0.190 \\ 
           & V &       -  &       -  &    0.800 &    1.277  &       -  &       -  &   28.313 &   26.735&       -  \\ 
\hline                                                                                                 
   3DMCPOL & I &    0.092 &    0.010 &    0.051 &    0.009  &    2.432 &    0.213 &    2.665 &    0.197&    0.499 \\ 
           & Q &    1.681 &    0.354 &    0.117 &    0.041  &    1.221 &    0.986 &    1.094 &    1.070&    2.574 \\ 
           & U &    2.129 &    0.275 &    0.108 &    0.061  &     -    &      -   &    1.264 &    1.198&   22.359 \\ 
           & V &       -  &       -  &    0.519 &    1.840  &     -    &      -   &   36.313 &   34.160&       -  \\ 
\hline                                                                                                 
    SPARTA & I &    0.088 &    0.011 &    0.051 &    0.027  &    0.183 &    0.198 &    0.213 &    0.143&    0.146 \\ 
           & Q &    0.367 &    0.055 &    0.120 &    0.041  &    2.256 &    1.725 &    2.050 &    2.011&    0.152 \\ 
           & U &    0.275 &    0.042 &    0.084 &    0.060  &      -   &      -   &    2.928 &    2.881&    0.231 \\ 
           & V &       -  &       -  &    0.639 &    2.607  &      -   &      -   &   67.027 &   64.972&       -  \\ 
\hline                                                                                                 
    SHDOM  & I &    0.044 &    0.068 &    0.111 &    0.077  &    3.383 &    0.077 &    3.640 &    0.075&    0.089 \\ 
           & Q &    0.034 &    0.233 &    0.274 &    0.051  &    8.735 &    9.311 &    0.841 &    0.805&    0.128 \\ 
           & U &    0.038 &    0.112 &    0.270 &    0.055  &       -  &       -  &    1.084 &    1.065&    0.195 \\ 
           & V &       -  &       -  &    0.567 &    1.818  &       -  &       -  &   28.745 &   27.062&       -  \\ 
\hline                                                                                                 
     Pstar & I &    0.017 &    0.009 &    0.100 &    0.025  &    0.336 &    0.081 &    0.159 &    0.075&    0.100 \\ 
           & Q &    0.024 &    0.036 &    0.289 &    0.028  &    0.809 &    0.775 &    0.827 &    0.811&    0.165 \\ 
           & U &    0.029 &    0.029 &    0.299 &    0.035  &       -  &       -  &    1.048 &    1.029&    0.273 \\ 
           & V &       -  &       -  &    0.808 &    1.278  &       -  &       -  &   28.302 &   26.732&       -  \\ 
\hline 

 \end{tabular}
 \caption{Relative root mean square differences  $\Delta_m$ in per cent between MYSTIC and
    IPOL, 3DMCPOL, SPARTA, SHDOM and Pstar for the single layer
    intercomparison cases.
    For the cloud cases the columns A5$^{\rm pp}_{\rm part}$ and
    A5$^{\rm al}_{\rm part}$ are added which include  $\Delta_m$
    calculated without the solar aureole region,  
    i.e. viewing angles up to 10\degree\ from the sun direction are taken
    out of the summation in Eq.~\ref{eq:rel_diff}. }
  \label{tab:a_results}
\end{table*}

\subsubsection{A1 -- Rayleigh scattering}

\begin{figure}[htbp]
  \centering
  \includegraphics[width=1.\hsize]{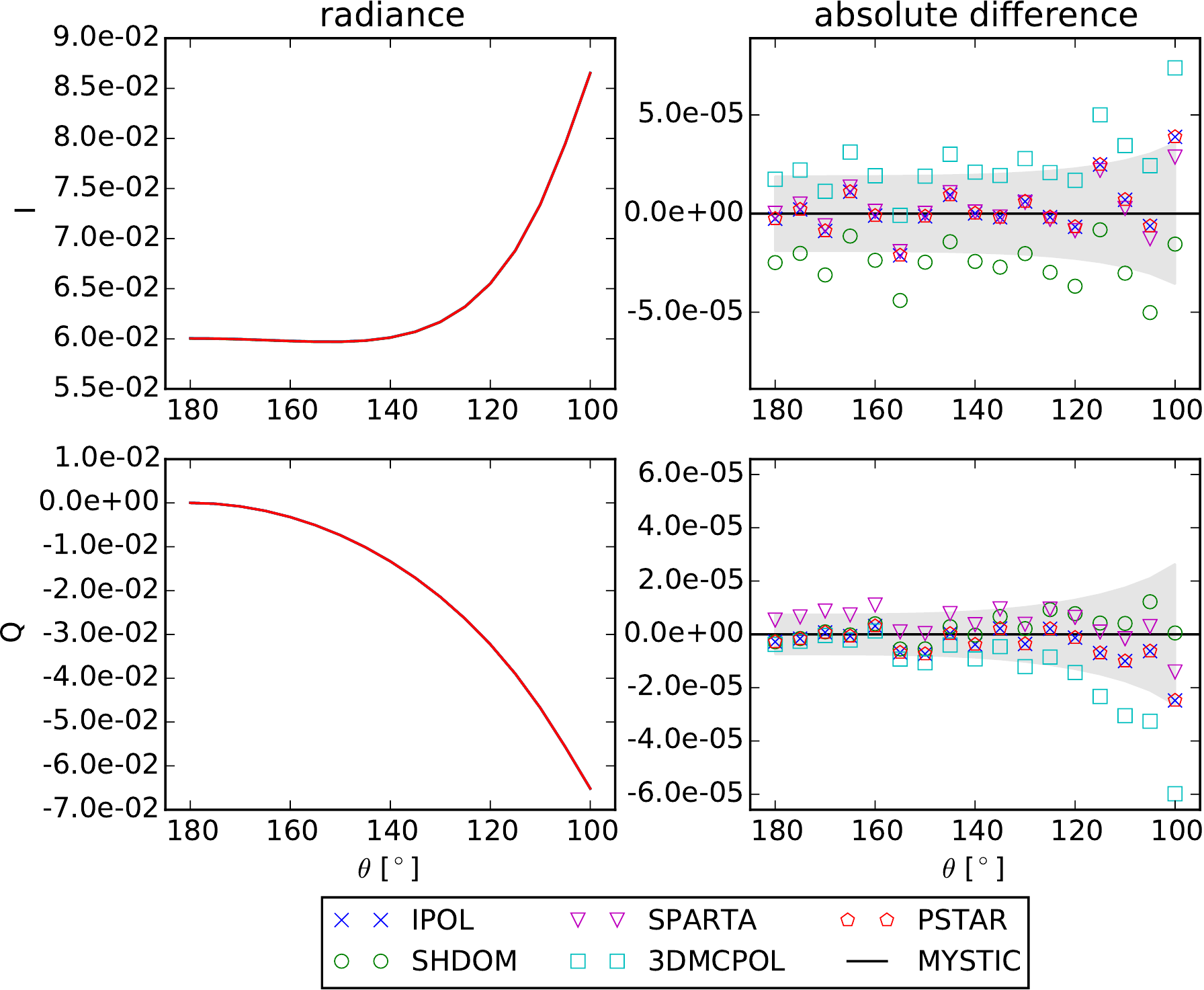}
  \caption{Test case A1, Rayleigh scattering layer, 
    depolarization factor 0, sun in the zenith, $\phi$=90\degree.  
    Left: Stokes vector at the top of layer. Right: Absolute difference between individual
    models and MYSTIC, the grey area corresponds to 2$\sigma$ of the
    Monte Carlo calculations.} 
  \label{fig:a1_results_1}
\end{figure}

The left plots in Fig.~\ref{fig:a1_results_1} show the Stokes vector calculated at the
top of the layer for down-looking directions. The sun is located in
the zenith, which means that $U$ is 0. 
The Rayleigh depolarization factor in this case is
0. The right plots show the absolute differences between the
models. The grey area corresponds to two standard deviations (2$\sigma$)
of the MYSTIC results, this means that with a probability of 95.4\%
the difference between the MYSTIC result and the true value lies in
the grey area. 

\label{sec:a1_results}

Looking at the left plots, we do not see differences between
the models, all lines are on top of each other. 
The right plots show that the differences between the models are three
orders of magnitude smaller than the radiance values. The level of
agreement between IPOL and Pstar is even better since the symbols of the
two models always lie on top of each other. This is not surprising
because the models use the same method. For IPOL, SPARTA and Pstar
the differences are centered about 0 and they are well in the 2$\sigma$
range, hence we may conclude that these models agree perfectly with
MYSTIC on a very high accuracy level. For $I$ the SHDOM results are
slightly smaller than MYSTIC whereas the 3DMCPOL results
are slightly larger. For $Q$, 3DMCPOL is slightly
smaller than MYSTIC. The difference plots show a similar progression for all
models, this is due to statistical noise of the MYSTIC results. The Monte Carlo
models SPARTA and 3DMCPOL use a technique to sample all directions
based on the same photon paths. In this case the statistical error is
the same for all viewing directions whereas for MYSTIC each direction
is calculated separately with an independent statistical error.

\begin{figure}[htbp]
  \centering
  \includegraphics[width=1.\hsize]{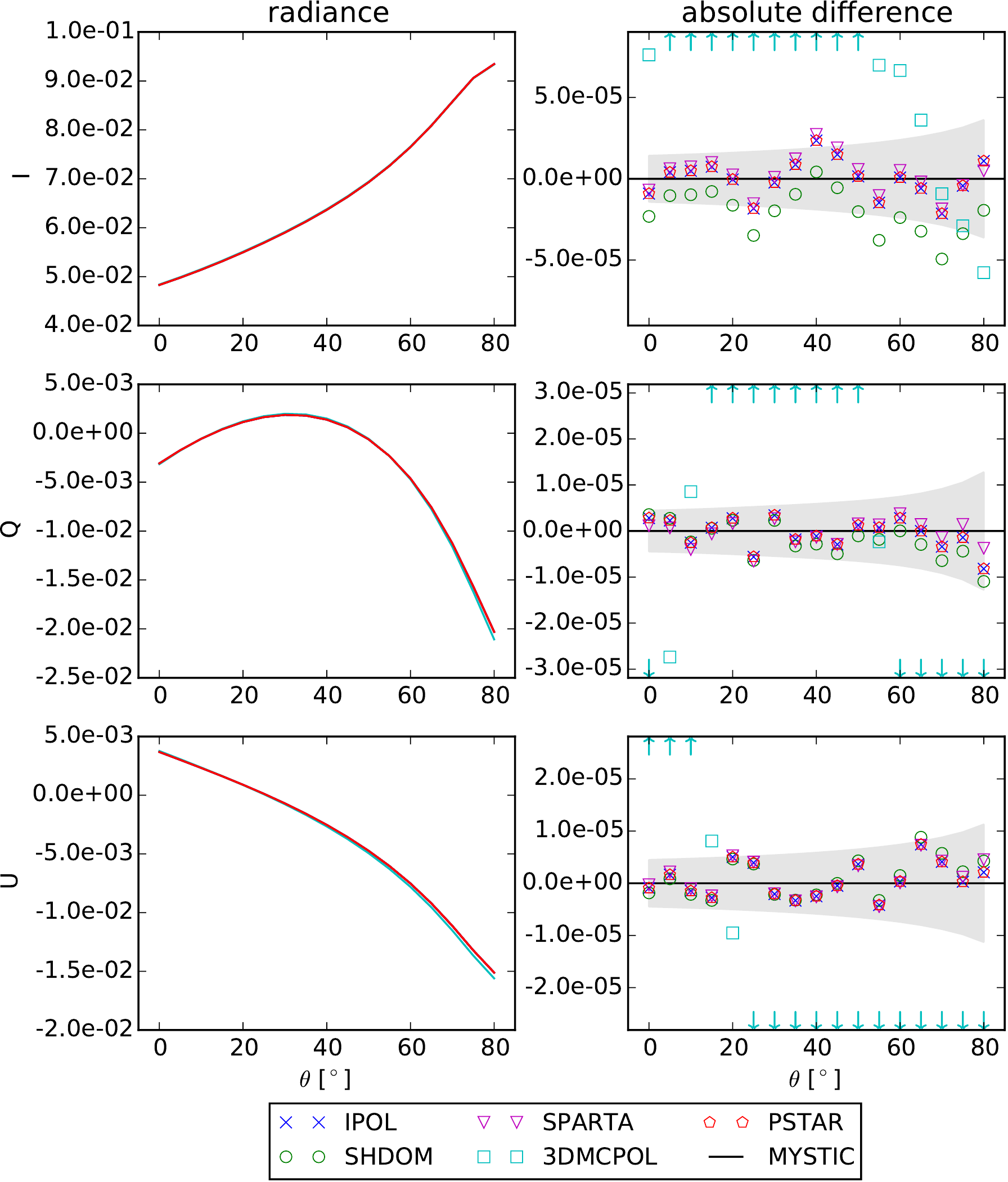}
  \caption{ Test case A1, Rayleigh scattering layer, 
    depolarization factor 0.1, sun at $(\theta_0=30\degree,
    \phi_0=65\degree)$, $\phi$=90\degree. 
    Left: Stokes vector at the surface. Right: Absolute difference between individual
    models and MYSTIC, the grey area corresponds to 2$\sigma$. 
    Arrows indicate out-of-range values.} 
  \label{fig:a1_results_2}
\end{figure}
Fig.~\ref{fig:a1_results_2} shows results for a Rayleigh scattering
matrix with depolarization factor 0.1. Here the solar zenith angle is
30\degree, hence $Q$ and $U$ are non-zero. In the left plots we see
that almost all model results are on top of each other, only the
3DMCPOL values are slightly below the results of the other models. The
maximum deviation is about 4\% for $Q$, the differences between MYSTIC
and 3DMCPOL are mostly out of range in the right plots. This indicates
that the depolarization factor in 3DMCPOL is not correctly implemented.
The depolarization factor bias
observed for 3DMCPOL was not corrected, consequently it has some
impacts on the results of the next sections.
The models IPOL and Pstar again agree perfectly, as in
Fig.~\ref{fig:a1_results_2} the symbols showing the differences to
MYSTIC lie on top of each other. The SPARTA results are very close to
IPOL and Pstar. The differences between IPOL, Pstar and SPARTA are
centered about 0 and well within the 2$\sigma$ range, hence they agree
perfectly with MYSTIC. For SHDOM the same is true for $Q$ and $U$,
whereas for $I$ SHDOM is systematically slightly smaller than MYSTIC.

Tab.~\ref{tab:a_results} shows that for test case A1,  $\Delta_m$ is smaller than
0.03\% for all Stokes components for the models IPOL, Pstar. Indeed the
numbers of $\Delta_m$ are exactly the same for IPOL and Pstar which
shows that the models use exactly the same method to solve the
radiative transfer equation for Rayleigh scattering. For SHDOM
$\Delta_m$ is smaller than 0.05\%, for SPARTA smaller than 0.4\% and for
3DMCPOL smaller than 2.1\%. 

\subsubsection{A2 -- Rayleigh atmosphere above Lambertian surface}
\label{sec:a2_results}
\begin{figure}[t]
  \centering
  \includegraphics[width=1.\hsize]{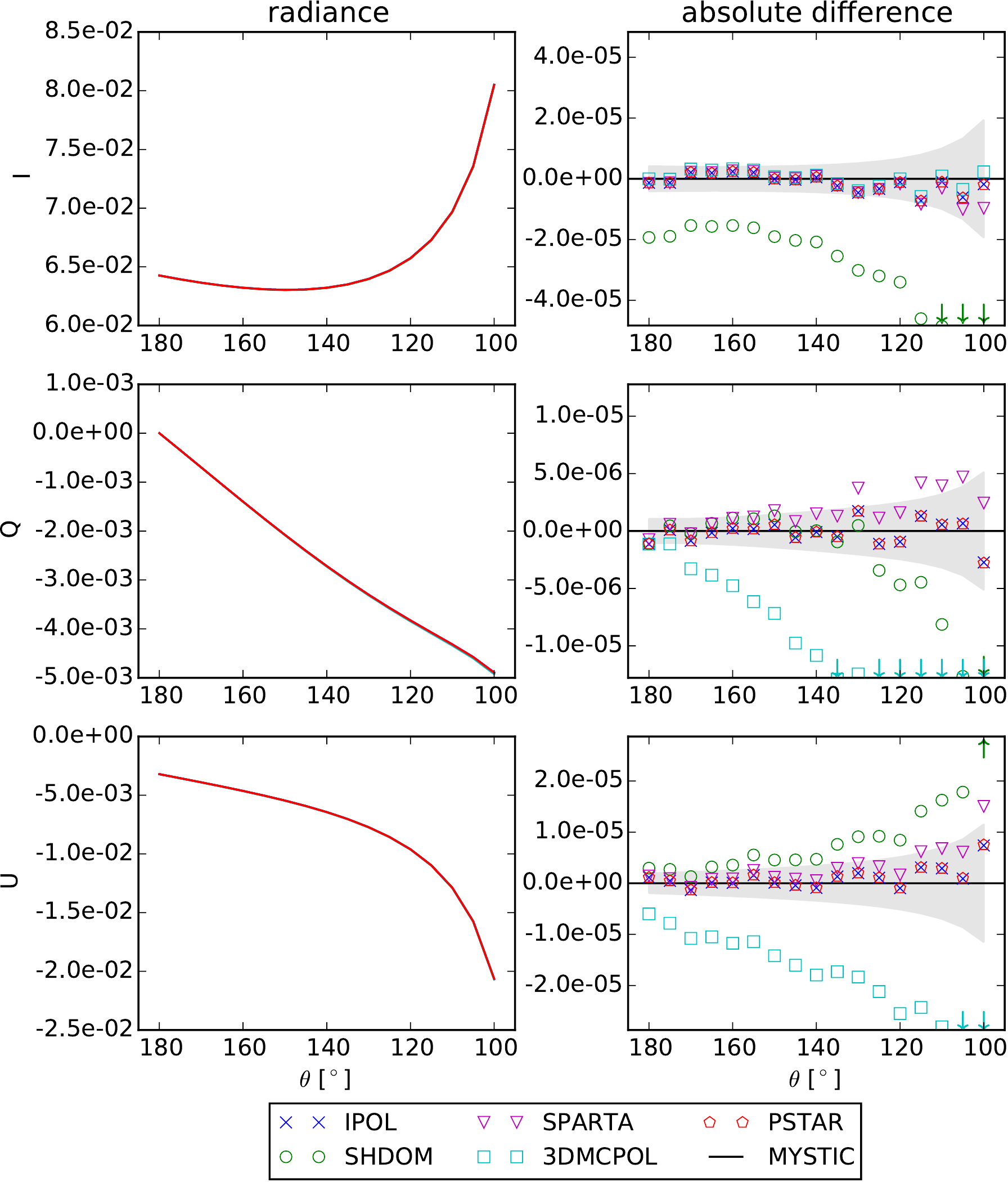}
  \caption{Test case A2, Rayleigh scattering layer (depolarization
    factor 0.03) above Lambertian
    surface with albedo 0.3, sun at $(\theta_0=50\degree,
    \phi_0=0\degree)$, $\phi$=45\degree. 
    Left: Stokes vector at top of atmosphere. 
    surface reflection. Right: Absolute difference between individual
    models and MYSTIC, the grey area corresponds to 2$\sigma$. Arrows
    indicate out-of-range values.} 
  \label{fig:a2_results}
\end{figure}
Fig.~\ref{fig:a2_results} shows results for the Rayleigh atmosphere 
above a Lambertian surface. In the radiance plots on the left side we
see that all lines are on top of each other. The difference plots show that
again, the models IPOL and Pstar agree, their symbols are exactly
on top of each other. The differences for IPOL, Pstar and SPARTA
scatter around 0 and they mostly lie within the 2$\sigma$ range, hence
these models agree perfectly within the expected accuracy of the
MYSTIC results. For $I$ the same is true for 3DMCPOL, whereas for $Q$
and $U$ the 3DMCPOL results are systematically smaller than
MYSTIC. The reason for this small, but systematic difference could be,
as for the case without surface, 
a slightly different implementation of the Rayleigh depolarization
factor which is 0.03 in this case.
The SHDOM results are slightly below MYSTIC for $I$ and $Q$ and
slightly above for $U$.   

Quantitatively we see in Tab.~\ref{tab:a_results} that the level of
agreement between MYSTIC and the models IPOL, Pstar and SPARTA is
$\Delta_m<$0.05\% and for
SHDOM and 3DMCPOL $\Delta_m<$0.35\%.

\subsubsection{A3 -- Spherical aerosol particles}
\label{sec:a3_results}
\begin{figure}[t]
  \centering
  \includegraphics[width=1.\hsize]{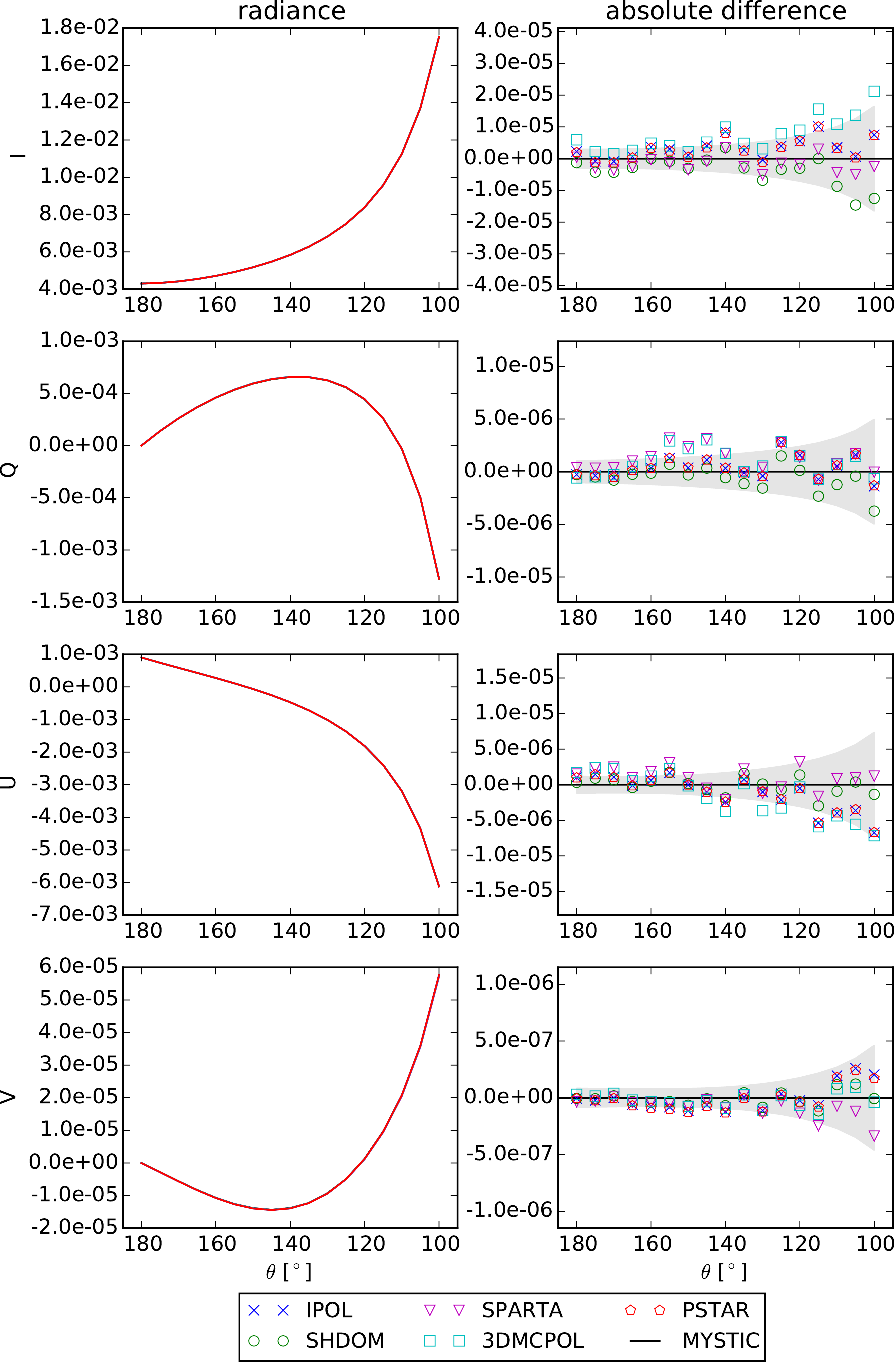}
  \caption{Test case A3, layer with spherical aerosol particles,
    optical thickness 0.2, sun
    at $(\theta_0=40\degree, \phi_0=0\degree)$, $\phi$=135\degree. 
    Left: Stokes vector at
    top of atmosphere.  Right: Absolute difference between individual
    models and MYSTIC, the grey area corresponds to 2$\sigma$.} 
  \label{fig:a3_results}
\end{figure}
Fig.~\ref{fig:a3_results} shows that all models agree very well for
the layer with the small spherical aerosol particles. The left side of
the plot shows that all models produce the full Stokes vector, also
the component for circular polarization $V$ very accurately, all curves
are here on top of each other. The difference plots on the right show
that the differences for all models lie in the 2$\sigma$ range. 
For $I$ the 3DMCPOL results seem systematically
larger than MYSTIC, although they are still in the 2$\sigma$ range.

The relative root mean square differences are $<$0.3\% for $I$,
$Q$ and $U$ and $<$0.8\% for $V$ (see Tab.~\ref{tab:a_results}).

\subsubsection{A4 -- Spheroidal aerosol particles}
\label{sec:a4_results}
\begin{figure}[t]
  \centering
  \includegraphics[width=1.\hsize]{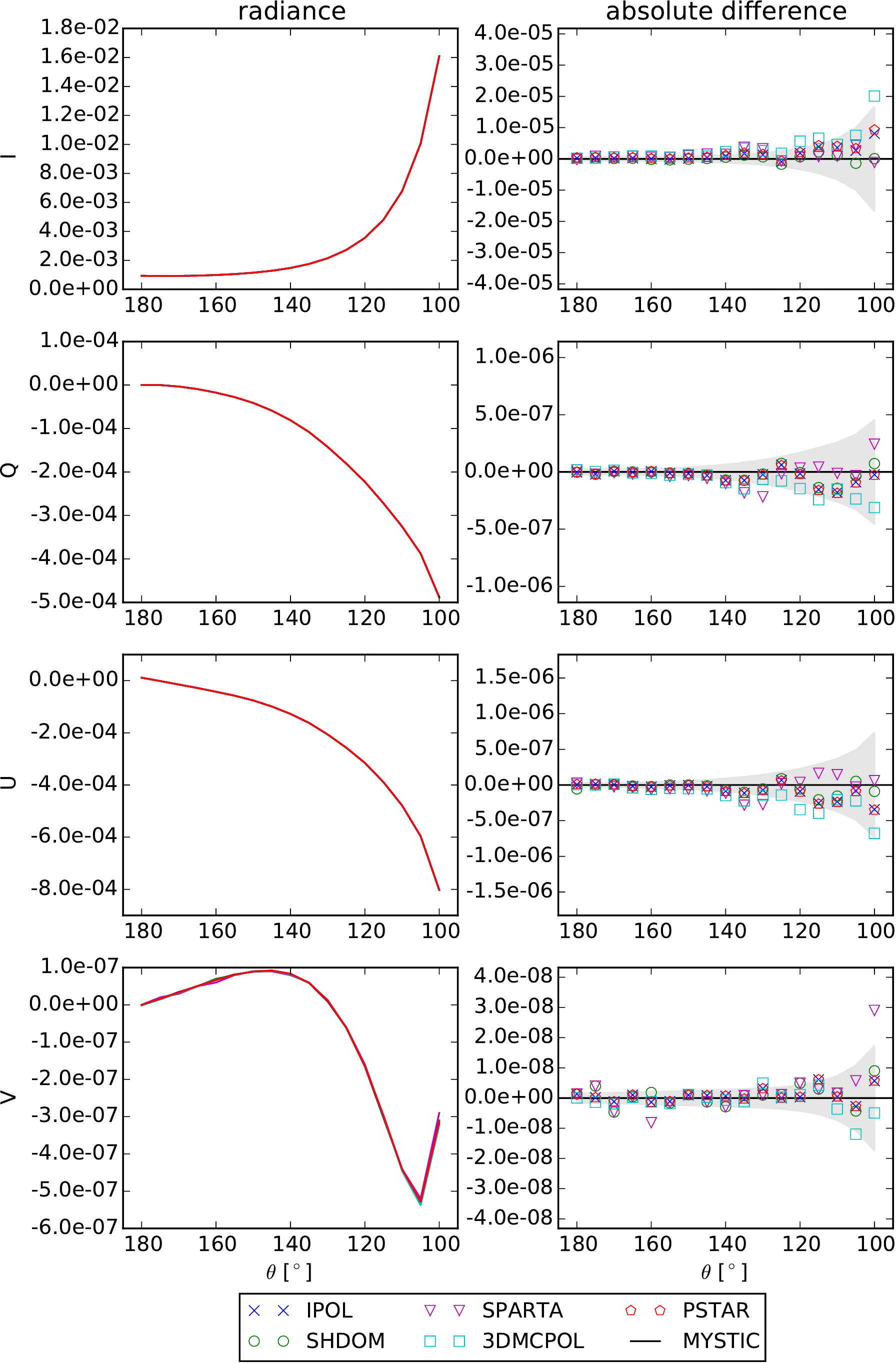}
  \caption{Test case A4, layer with spheroidal aerosol particles, optical thickness 0.2, sun
    at $(\theta_0=40\degree, \phi_0=0\degree)$, $\phi$=45\degree.
    Left: Stokes vector at the top of the atmosphere. Right: Absolute difference between individual
    models and MYSTIC, the grey area corresponds to 2$\sigma$.}
  \label{fig:a4_results1}
\end{figure}
The left plots in Fig.~\ref{fig:a4_results1} show the Stokes vector
at the top of the atmosphere, again all lines are on top of each other.
The absolute
differences between MYSTIC and all other models are in the 2$\sigma$
range for all Stokes components. 
The scattering phase function for the particles considered here shows
strong forward scattering. 
\begin{figure}[t]
  \centering
  \includegraphics[width=1.\hsize]{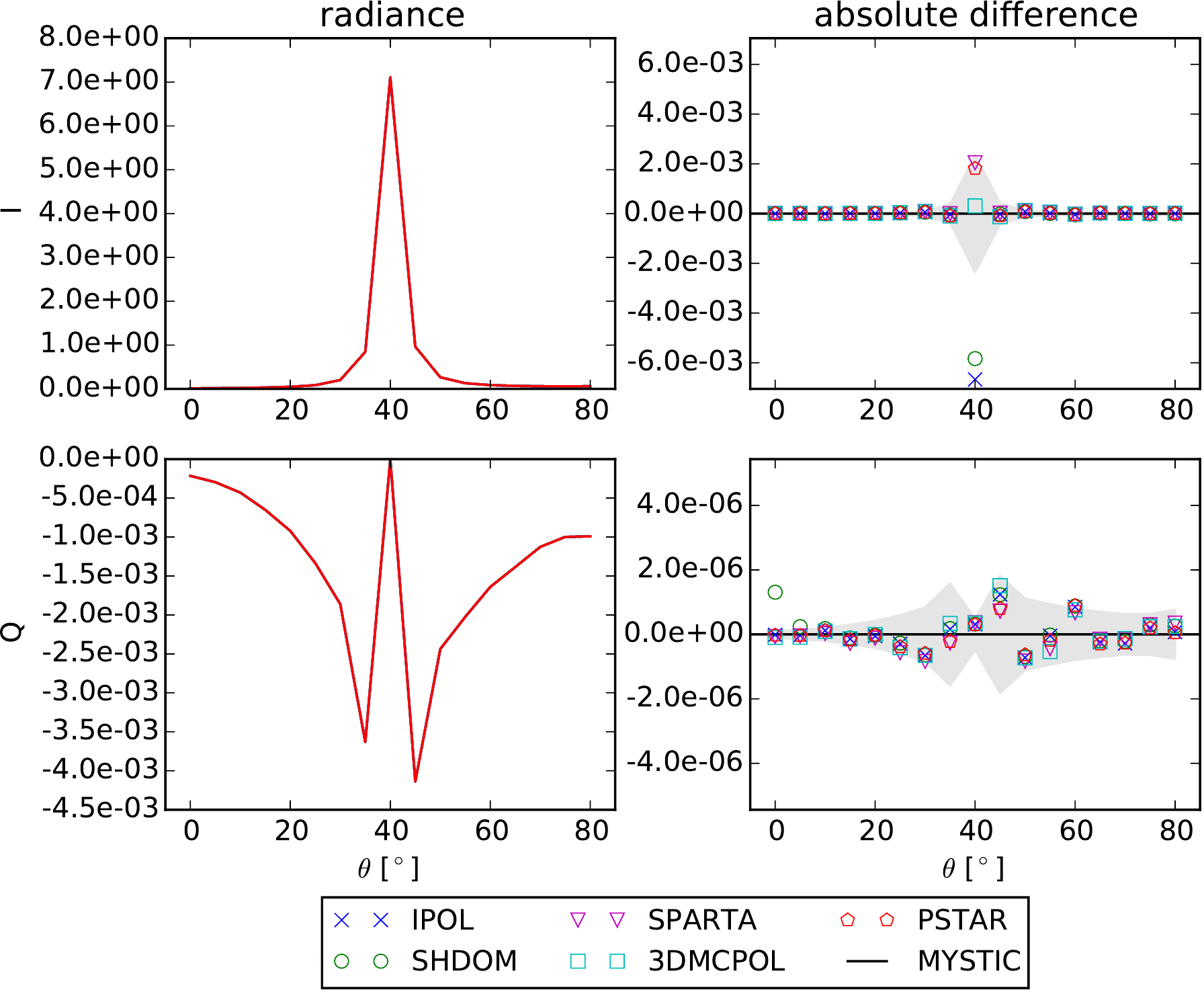}
  \caption{Test case A4, settings as in Fig.~\ref{fig:a4_results1} but
    for $\phi$=0\degree.  
    Left: Stokes vector at the surface.
    Right: Absolute difference between individual
    models and MYSTIC, the grey area corresponds to 2$\sigma$.} 
  \label{fig:a4_results2}
\end{figure}
Fig.~\ref{fig:a4_results2} shows Stokes
vector at the surface for a viewing azimuth of 0\degree, for which the
$U$ and $V$ are 0. This geometry
includes the sun direction and the forward scattering peak. The left
plots show, that also the forward scattering peak is calculated
accurately by all models. 
The models 3DMCPOL, Pstar and SPARTA agree to MYSTIC within the 2$\sigma$
range, even in exact forward scattering directions,
where SHDOM and IPOL are a little lower (0.1\%). 
For $Q$ all models agree with MYSTIC in the expected
2$\sigma$ range. 

Tab.~\ref{tab:a_results} shows that the relative root mean square
difference between MYSTIC and all models is
0.09\% or better for $I$, $Q$, and $U$. 
For $V$ the differences are
of the order of 1--3\%. The absolute value of $V$ is of the order
of $10^{-7}$ and the statistical uncertainty
of the Monte Carlo results for this small radiances is about 1--3\%.

\subsubsection{A5 -- Liquid water cloud}
\label{sec:a5_results}
\begin{figure}[htbp]
  \centering
  \includegraphics[width=1.\hsize]{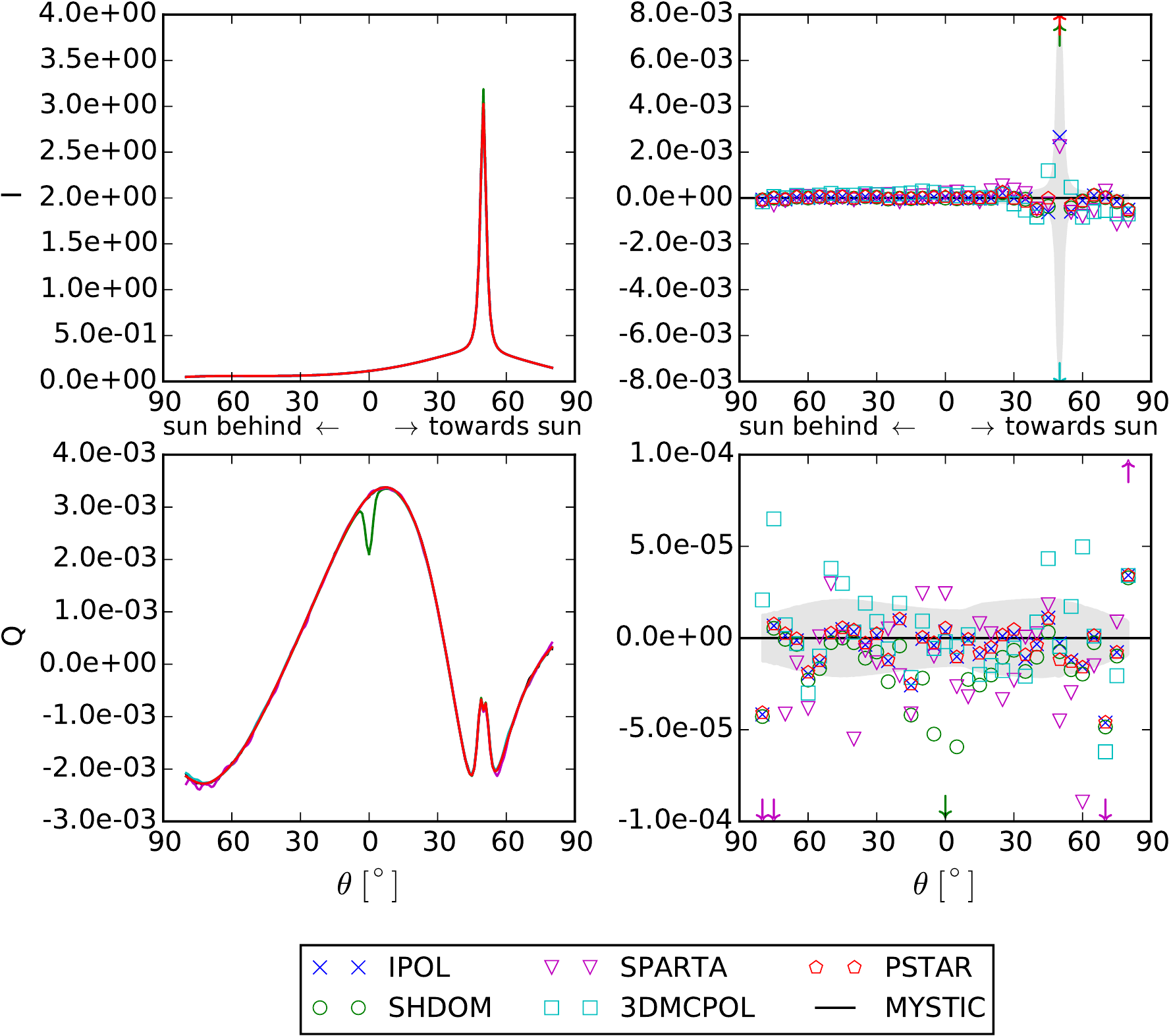}
  \caption{Test case A5, cloud layer with optical thickness of 5,
    effective radius of 10, principal plane, sun at $(\theta_0=50\degree, \phi_0=0\degree)$.
    Left: Stokes vector at the surface. Right: Absolute difference between individual
    models and MYSTIC, the grey area corresponds to 2$\sigma$.
    A marker is plotted only at every 5\degree, although the
    simulations were done in 1\degree\ steps.
    Arrows indicate out-of-range values.} 
  \label{fig:a5_results1}
\end{figure}
Fig.~\ref{fig:a5_results1} shows the principal plane. In the
region of the forward scattering peak the models IPOL and SPARTA agree
to  MYSTIC within
the expected accuracy (2$\sigma$ of MYSTIC calculation).
For 3DMCPOL the value of the forward
scattering peak in total intensity $I$ about 3\% smaller whereas
for Pstar and SHDOM it is larger (about 0.5\% and 6\% respectively). 
SHDOM has an artefact in the second
Stokes component Q around 0$^{\circ}$ viewing zenith angle.  Further
tests have shown that this artefact around 0$^{\circ}$ and 180$^{\circ}$
viewing zenith angles occurs only with highly peaked phase functions and
are largest for moderate optical depths (there is no artefact for single
scattering).  The width of the artefact decreases with higher SHDOM
angular resolution.  The artefact is believed to be a result of a
deficiency in the delta-M formulation for polarization.
\begin{figure}[t]
  \centering
  \includegraphics[width=1.\hsize]{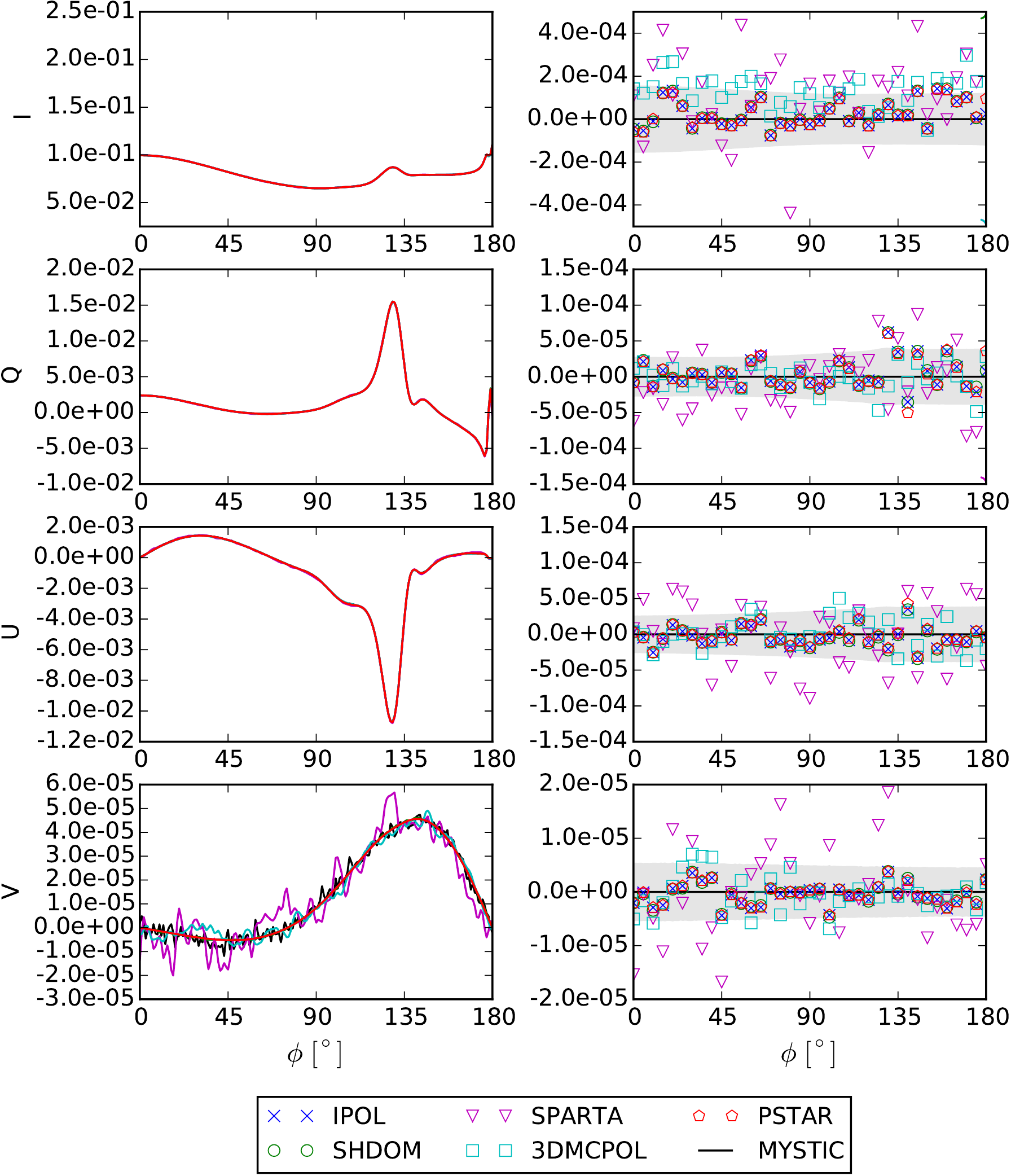}
  \caption{Test case A5, settings as in Fig:~\ref{fig:a5_results1} but
    for ``almucantar plane'', viewing zenith angle of 50\degree. 
    Left: Stokes vector at the top of the layer. Right: Absolute difference between individual
    models and MYSTIC, the grey area corresponds to 2$\sigma$.
    A marker is plotted only at every 5\degree, although the
    simulations were done in 1\degree\ steps.} 
  \label{fig:a5_results2}
\end{figure}
Fig.~\ref{fig:a5_results2} shows the reflected radiance in the
``almucantar'' plane, i.e. the viewing zenith angle is constant and
corresponds to the solar zenith angle of 50\degree. 
Here we see clearly that the standard
deviation of 3DMCPOL and SPARTA is a little higher than for MYSTIC,
therefore several points are outside the 2$\sigma$ range.
For SPARTA this is not surprising because it does not use any variance
reduction techniques for highly asymmetric scattering phase
functions. For the very small
Stokes component $V$, all Monte Carlo models are quite noisy, which can
be seen in the radiance plot for $V$.  Using more photons or including
better variance reduction method could decrease the noise. 
Within  the Monte Carlo noise the models 
agree perfectly.

Tab.~\ref{tab:a_results} shows that the smallest relative root mean square
difference is found for IPOL and Pstar, with  $\Delta_m <$0.4\% for
$I$ and values about 1\%
for $Q$ and $U$. For $V$, $\Delta_m$ is much larger, about 30\%. The
reason is the large noise in the MYSTIC calculations. Except for $V$, a
good agreement (within the range of 0.2\%-3\%) is found for the models
3DMCPOL and SPARTA. For SHDOM, $\Delta_m$ for $Q$ is dominated by the
artefact mentioned before. Without the specific directions (forward
scattering and 0\degree\ viewing zenith angle),
SHDOM also agrees perfectly to all other
models (see column A5$^{al}_{part}$ in Tab.~\ref{tab:a_results}).

\subsubsection{A6 -- Rayleigh atmosphere above ocean surface}
\label{sec:a6_results}
\begin{figure}[t]
  \centering
  \includegraphics[width=1.\hsize]{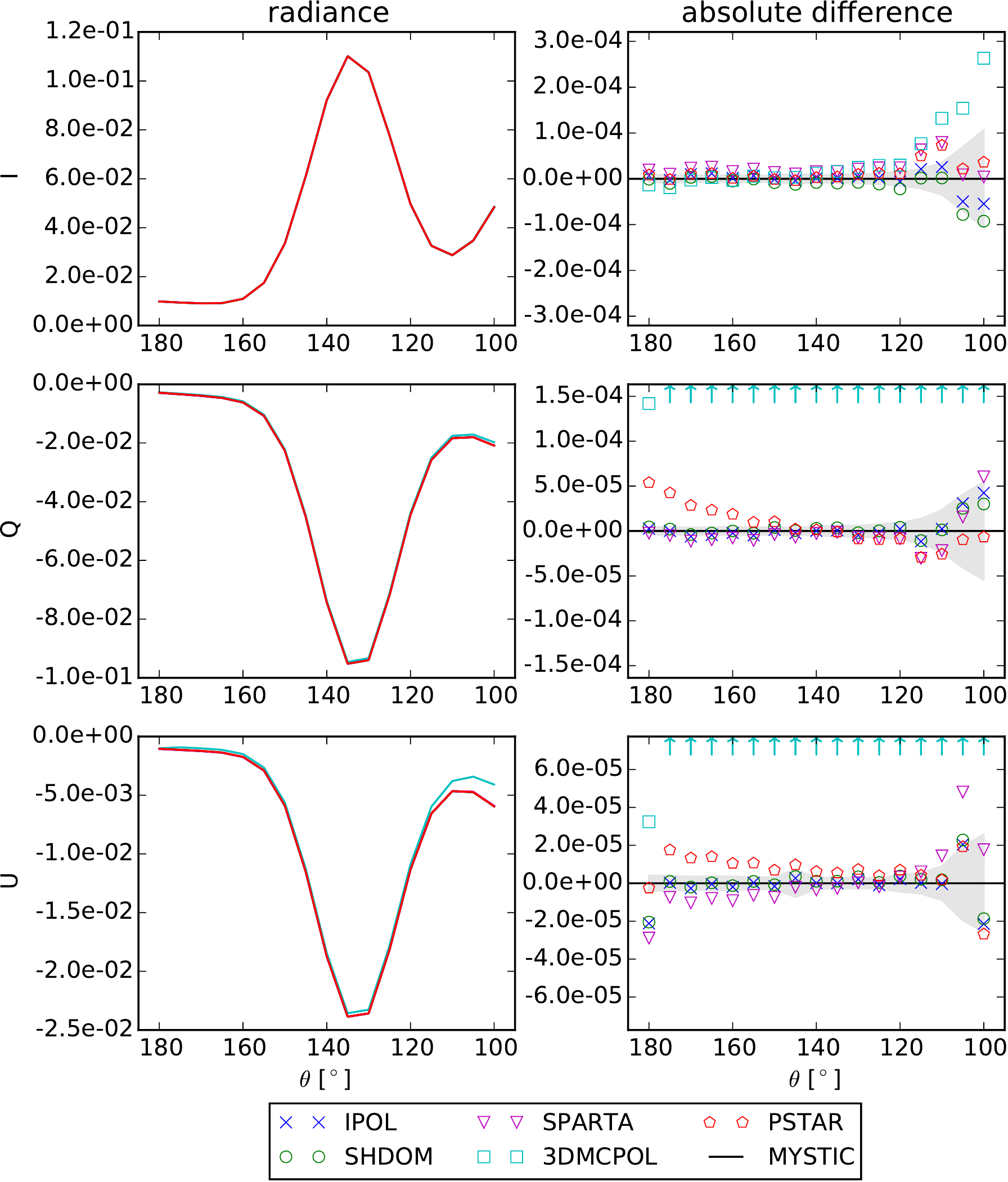}
  \caption{Test case A6, Rayleigh scattering layer above ocean surface
    with windspeed of 2~m/s,
    sun at $(\theta_0=45\degree, \phi_0=0\degree)$, $\phi$=10\degree. 
    Left: Stokes vector at top of atmosphere.
    Right: Absolute difference between individual
    models and MYSTIC, the grey area corresponds to 2$\sigma$.
    Arrows indicate out-of-range values.} 
  \label{fig:a6_results}
\end{figure}
Fig.~\ref{fig:a6_results} shows the reflected Stokes vector for the
Rayleigh scattering layer above the ocean surface. All models produce
the sunglint and the models MYSTIC, IPOL, SHDOM and SPARTA agree very well.
For Pstar, we see a small bias for all Stokes components in the shown
geometry. The 3DMCPOL results show larger deviations
for all Stokes components due to an error in the code which has not
been discovered so far.

The numbers in Tab.~\ref{tab:a_results} show that the models IPOL,
SPARTA, Pstar and SHDOM agree very well to MYSTIC with
$\Delta_m<\sim$0.3\%. The 3DMCPOL results are inaccurate especially for $U$. 

\subsection{Test cases with realistic atmospheric profiles}

For the multi-layer test cases, the radiation fields have been
calculated at the surface, at the top of the atmosphere and inside the
atmosphere at an altitude of 1~km. However, at present time
only MYSTIC, SHDOM and
Pstar are capable to calculate the radiance field inside the
atmosphere. 
The relative root mean square differences for the multi layer test cases are
listed in Table~\ref{tab:b_results}, the details are discussed in the
following sections.

 \begin{table*}[t!]
  \centering
  \begin{tabular}{|l l|c|c|c|c|c|c|}
    \hline
    model name &  & B1 & B2 & B3 & B3$_{\rm part}$ & B4 & B4$_{\rm part}$ \\ \hline \hline 
    IPOL & I &    0.016 &    0.012 &    0.060 &    0.014 &    0.111 &    0.091 \\ 
         & Q &    0.024 &    0.023 &    0.049 &    0.034 &    0.575 &    0.472 \\ 
         & U &    0.019 &    0.019 &    0.043 &    0.026 &    0.601 &    0.478 \\ 
         & V &      - &      - &    1.489 &    1.073 &   19.377 &   15.091 \\ 
\hline 
   3DMCPOL & I &    0.033 &    1.387 &    0.040 &    0.064 &    1.152 &    0.727 \\ 
           & Q &    0.558 &    0.975 &    1.296 &    1.261 &   27.614 &   14.805 \\ 
           & U &    0.505 &    0.517 &    1.037 &    0.979 &    5.965 &    5.878 \\ 
           & V &       -  &       -  &    4.347 &    3.833 &   94.928 &   62.132 \\ 
\hline 
    SPARTA & I &    0.020 &    0.013 &    0.055 &    0.026 &    0.344 &    0.326 \\ 
           & Q &    0.030 &    0.029 &    0.071 &    0.045 &    3.710 &    2.699 \\ 
           & U &    0.023 &    0.023 &    0.064 &    0.036 &    4.368 &    2.856 \\ 
           & V &       -  &       -  &    1.982 &    1.439 &  182.181 &   88.770 \\ 
\hline 
    SHDOM  & I &    0.052 &    0.054 &    0.426 &    0.069 &    1.059 &    0.109 \\ 
           & Q &    0.068 &    0.071 &    0.148 &    0.085 &    2.377 &    1.654 \\ 
           & U &    0.040 &    0.054 &    0.136 &    0.057 &    3.846 &    2.153 \\ 
           & V &       -  &       -  &    2.567 &    1.548 &   23.672 &   19.700 \\ 
\hline 
     Pstar & I &    0.017 &    0.013 &    0.154 &    0.017 &   34.947 &    0.182 \\ 
           & Q &    0.025 &    0.026 &    0.052 &    0.039 &    0.644 &    0.579 \\ 
           & U &    0.020 &    0.021 &    0.047 &    0.031 &    2.583 &    2.661 \\ 
           & V &       -  &       -  &    1.553 &    1.127 &   23.695 &   19.730 \\ 
\hline

  \end{tabular}
  \caption{Relative root mean square differences $\Delta_m$ in per cent between MYSTIC and
    IPOL, 3DMCPOL, SPARTA, SHDOM and Pstar for the multi-layer
    intercomparison cases. For  B3$_{\rm part}$ and B4$_{\rm part}$ the
    solar aureole region is taken out of the calculation of $\Delta_m$,
  i.e. viewing angles up to 10\degree\ from the sun direction are taken
out of the summation in Eq.~\ref{eq:rel_diff}.}
  \label{tab:b_results}
\end{table*} 

\subsubsection{B1 -- Rayleigh scattering for a standard atmosphere}
\label{sec:b1_results}
\begin{figure}[htbp]
  \centering
  \includegraphics[width=1.\hsize]{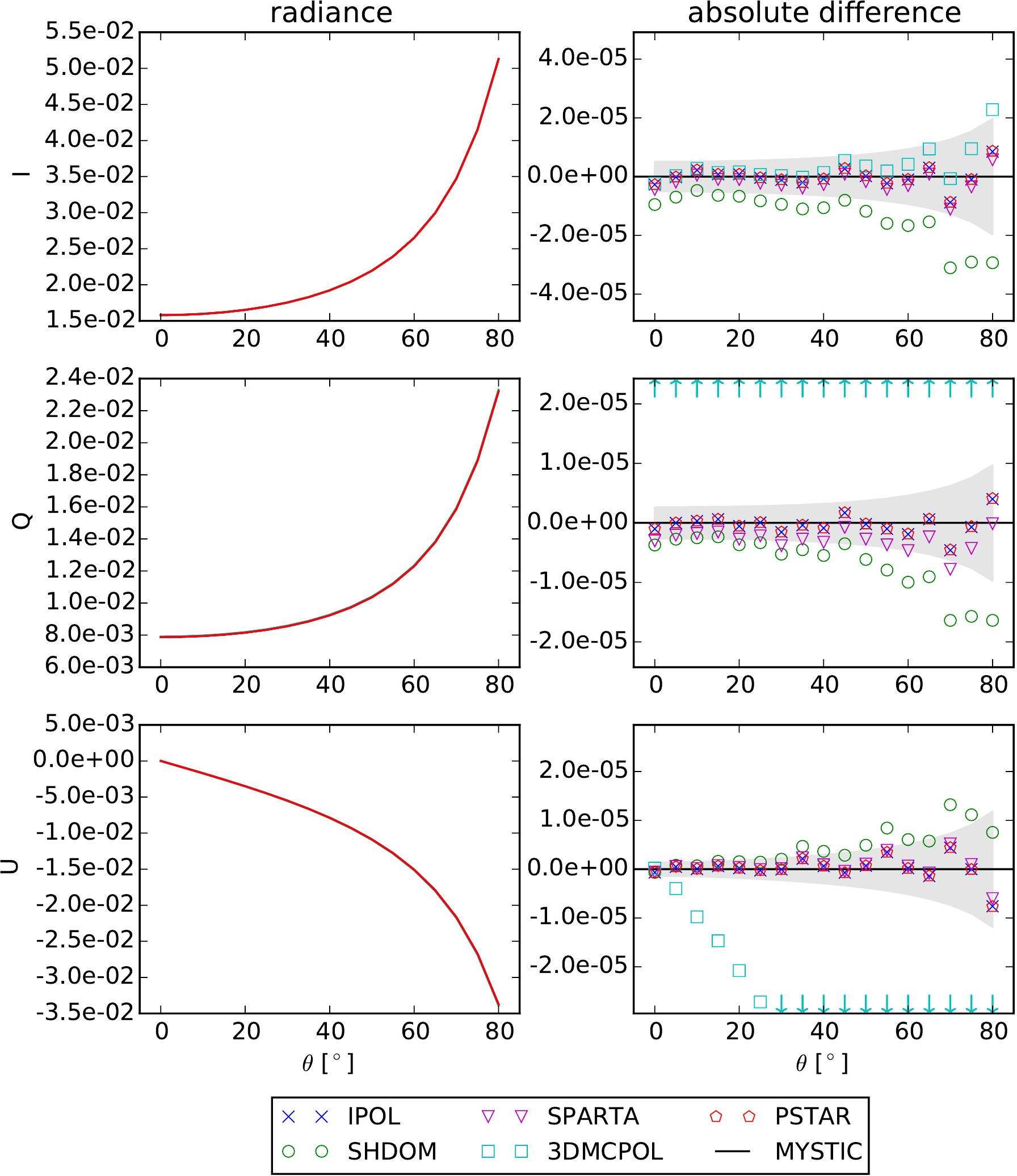}
  \caption{Test case B1, US-standard atmosphere, 450~nm, no absorption, sun at $(\theta_0=60\degree, \phi_0=0\degree)$, $\phi$=90\degree. Left: Stokes vector at the surface. Right: Absolute difference between individual
    models and MYSTIC, the grey area corresponds to 2$\sigma$.
    Arrows indicate out-of-range values.} 
  \label{fig:b1_results}
\end{figure}
Fig.~\ref{fig:b1_results} shows the results for a multi-layer
atmosphere with pure Rayleigh scattering. 
As for the 1-layer case, we find a very good agreement between all
models for pure Rayleigh scattering. The models IPOL, SPARTA and Pstar
agree among each other. SHDOM is slightly smaller
for $I$ and $Q$ and slightly larger for $U$. 3DMCPOL shows small but
systematic deviations, which might again be due to a different
implementation of the Rayleigh depolarization factor, which was set to
0.03 in this test case.

Tab.~\ref{tab:b_results} shows that relative root mean square
deviations are mostly smaller than 0.05\% between MYSTIC and IPOL,
SPARTA, SHDOM, and Pstar respectively. For 3DMCPOL,  $\Delta_m$ is
0.03\% for $I$ and about 0.5\% for $Q$ and $U$.

\subsubsection{B2 -- Rayleigh scattering and absorption for a standard atmosphere}
\label{sec:b2_results}
\begin{figure}[htbp]
  \centering
  \includegraphics[width=1.\hsize]{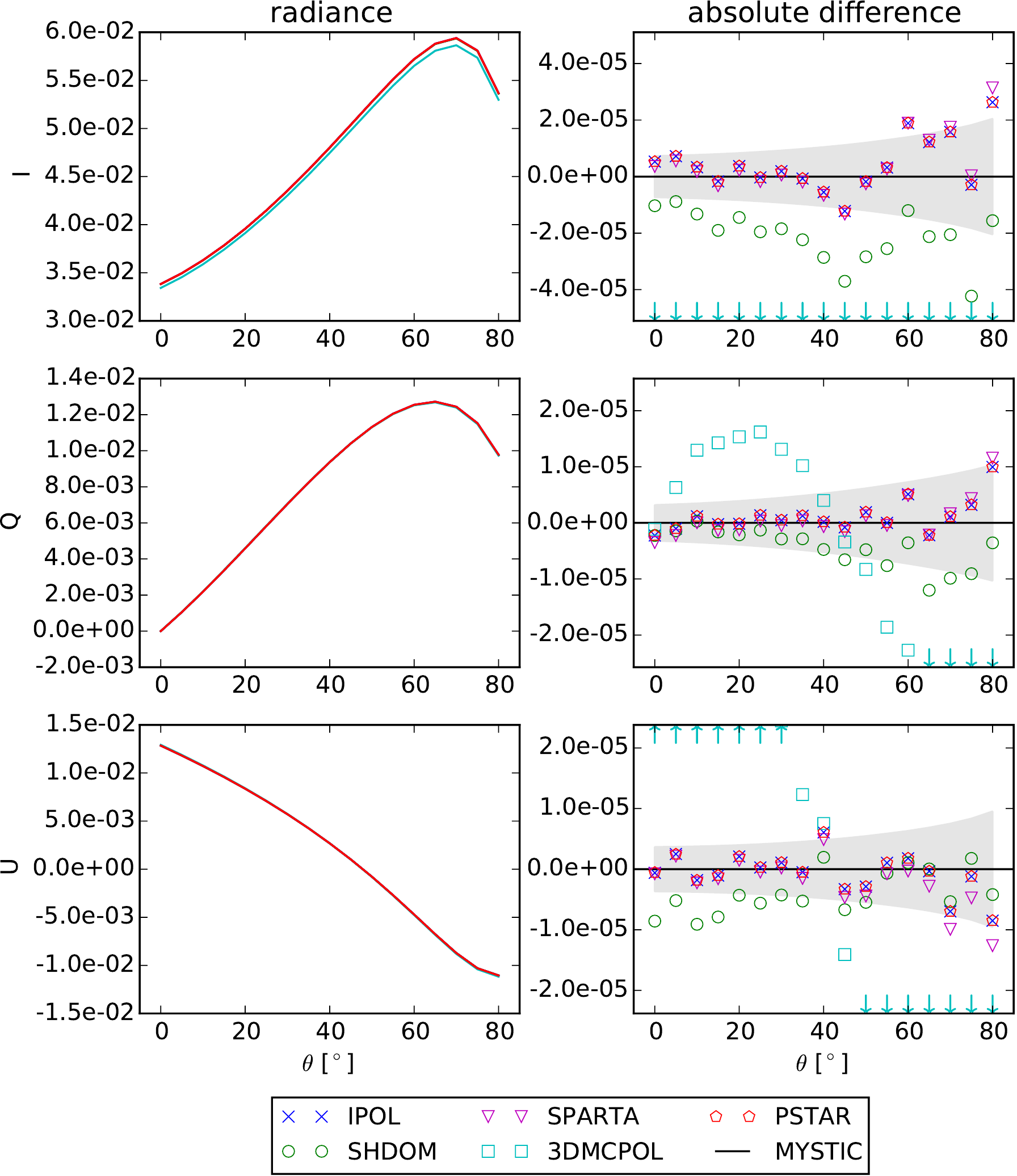}
  \caption{Test case B2, US-standard atmosphere, scattering an
    absorption, 325~nm, sun at  $(\theta_0=60\degree,
    \phi_0=0\degree)$, $\phi$=45\degree. 
    Left: Stokes vector at the surface.
    Right: Absolute difference between individual
    models and MYSTIC, the grey area corresponds to 2$\sigma$.
    Arrows indicate out-of-range values.} 
  \label{fig:b2_results}
\end{figure}
Fig.~\ref{fig:b2_results} shows the results for the US-standard
atmosphere simulated at 325~nm, where absorption has been included. 
The models IPOL, SPARTA and Pstar agree to MYSTIC within two standard
deviations. The SHDOM results show a small bias, for $I$ they are
slightly smaller than the MYSTIC results. The 3DMCPOL results differ by
more than 1\%.

In Tab.~\ref{tab:b_results} we see that relative root mean square
deviations are as for test case B1 
mostly smaller than 0.05\% between MYSTIC and IPOL,
SPARTA, SHDOM, and Pstar respectively. For 3DMCPOL,  $\Delta_m$ is
in the range of 0.5\%--1.5\% for all Stokes components.

\subsubsection{B3 -- Aerosol profile and standard atmosphere}
\label{sec:b3_results}

\begin{figure}[t]
  \centering
  \includegraphics[width=1.\hsize]{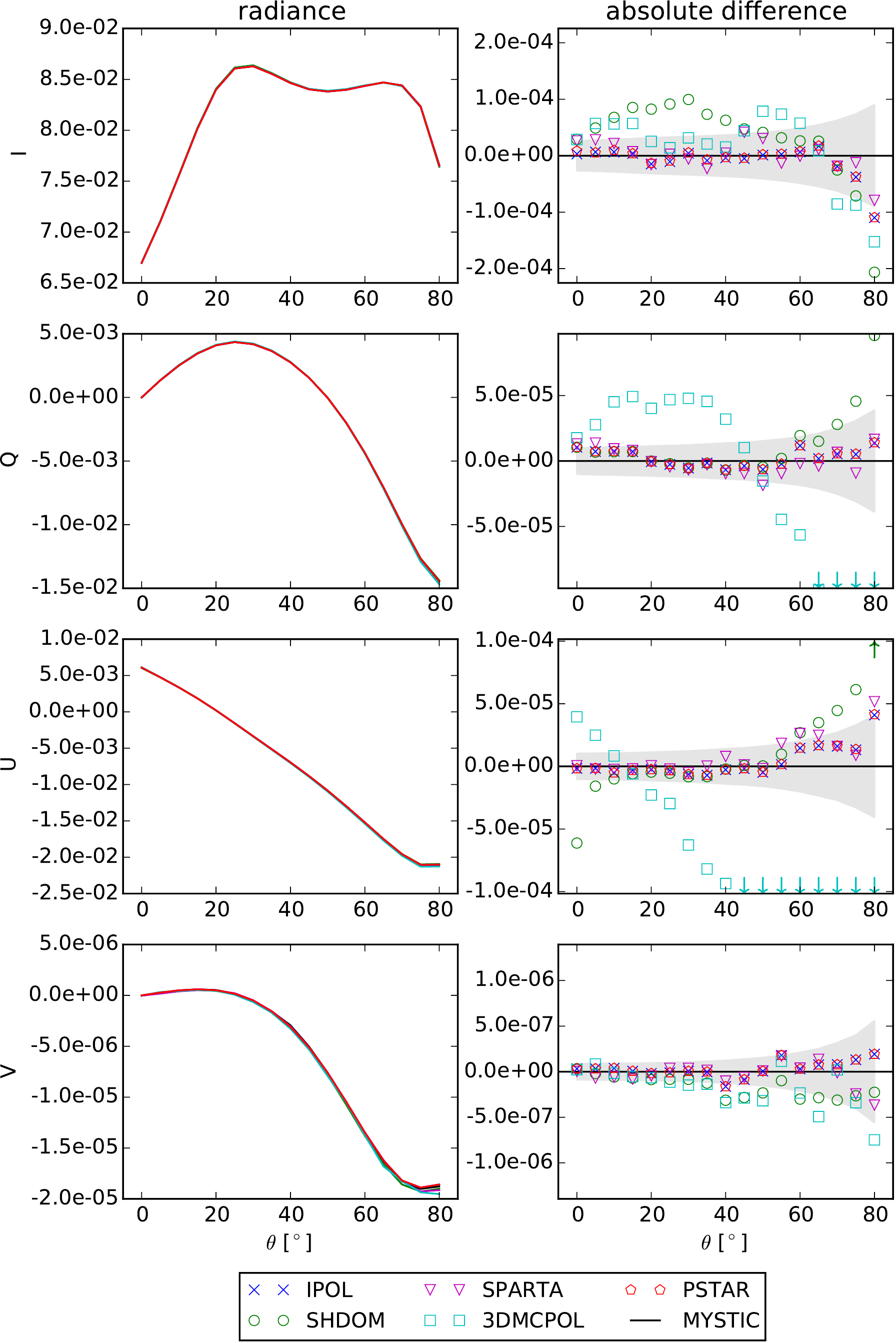}
  \caption{Test case B3, US-standard atmosphere, 350~nm, aerosol profile,
    spheroidal particles, sun at  $(\theta_0=30\degree,
    \phi_0=0\degree)$, $\phi$=45\degree. Left: Stokes vector at the
    surface. Right: Absolute difference between individual
    models and MYSTIC, the grey area corresponds to 2$\sigma$.
    Arrows indicate out-of-range values.} 
  \label{fig:b3_results}
\end{figure}
For the standard atmosphere including a realistic aerosol profile
again all models agree very well. Fig.~\ref{fig:b3_results} shows that
deviations between MYSTIC and
IPOL, SPARTA and Pstar are again within the expected
uncertainty for most angles. There are some tiny deviations for
SHDOM. 3DMCPOL again shows small systematic deviations, especially for $Q$
and $U$. 

The relative root mean square deviations from MYSTIC
(Tab.~\ref{tab:b_results}) are smaller than
0.07\% for $I$, $Q$ and $U$ for the models IPOL and SPARTA. For Pstar and
SHDOM, the differences are slightly larger (up to 0.4\%). 
$\Delta_m$ is about 1.5\% for $V$ for IPOL and
Pstar. This larger deviation is due to an increased relative standard
deviation of MYSTIC, because $V$ is four orders of magnitude smaller
than $Q$ and $U$. 
For 3DMCPOL, $\Delta_m\sim$0.04\% for $I$, of the order of 1\% for
$Q$ and $U$ and 4\% for $V$.

\subsubsection{B4 -- Cloud above ocean surface}
\label{sec:b4_results}

\begin{figure}[t]
  \centering
  \includegraphics[width=1.\hsize]{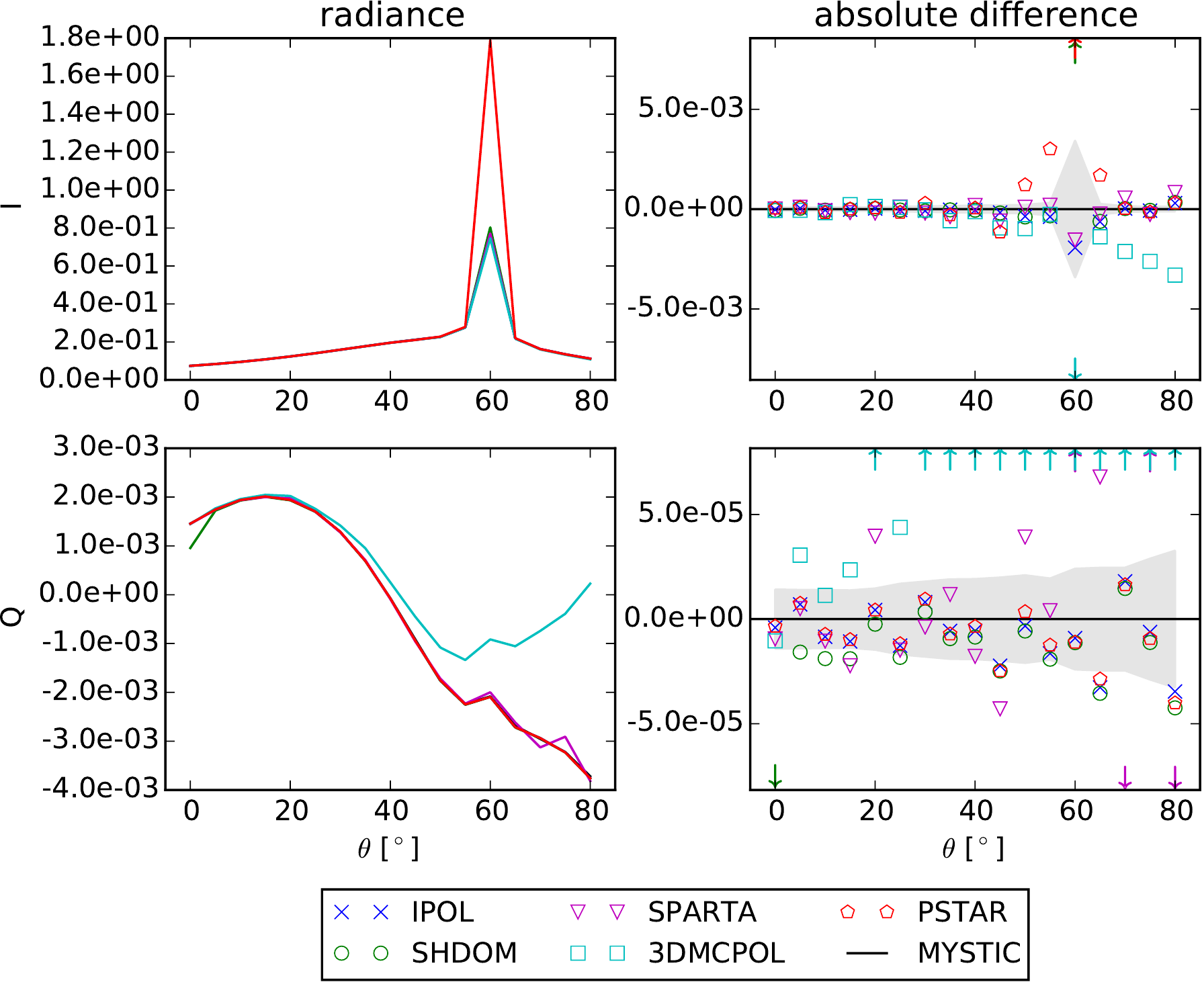}
  \caption{Test case B4, US-standard atmosphere, 800~nm, ocean
    surface, cloud layer with optical thickness of 5,
    sun at  $(\theta_0=60\degree,
    \phi_0=0\degree)$, $\phi$=0\degree. 
    Left: Stokes vector at the surface. 
    Right: Absolute difference between individual
    models and MYSTIC, the grey area corresponds to 2$\sigma$.
    Arrows indicate out-of-range values.} 
  \label{fig:b4_results1}
\end{figure}
\begin{figure}[t]
  \centering
  \includegraphics[width=1.\hsize]{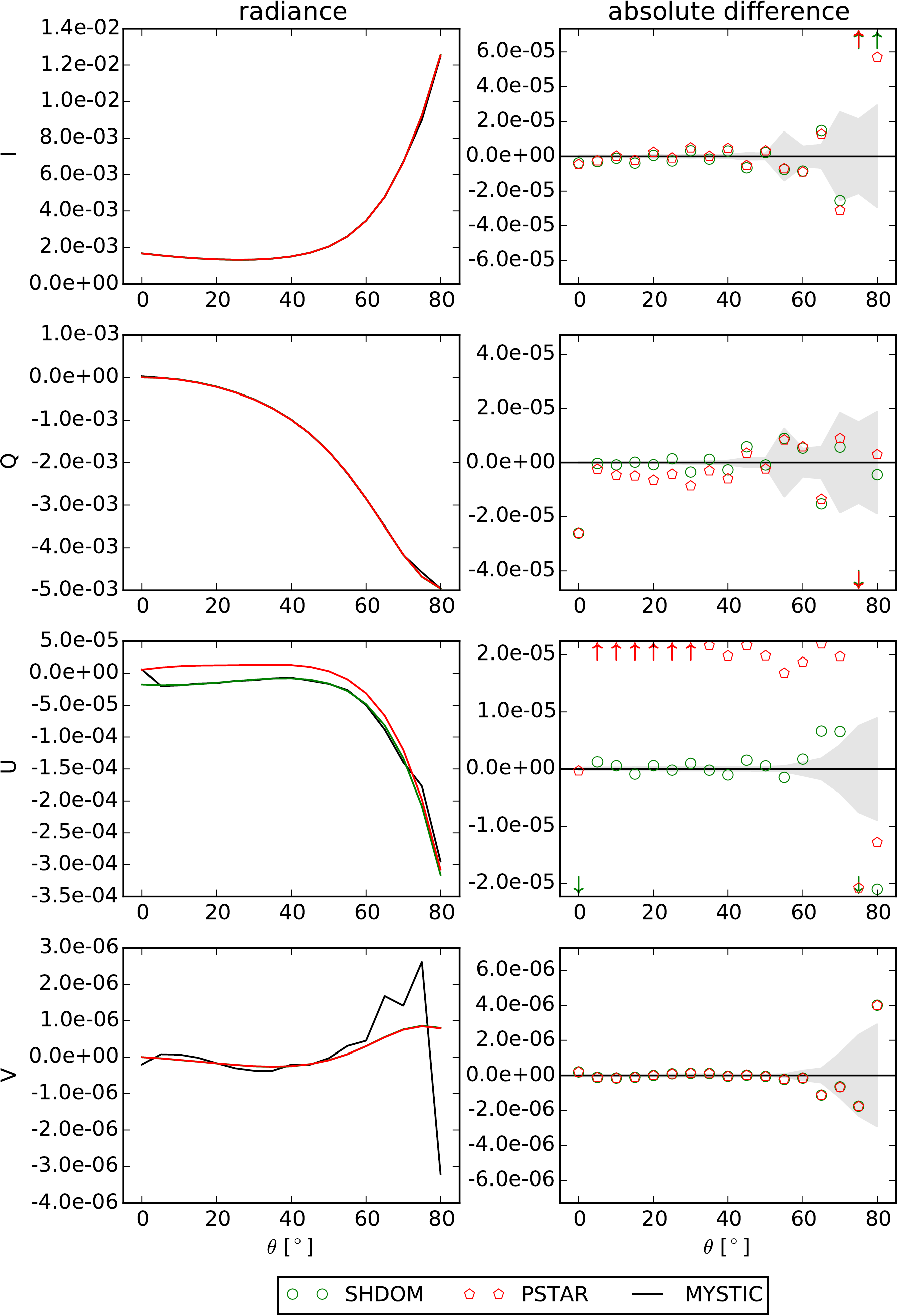}
  \caption{Test case B4, settings as Fig.~\ref{fig:b4_results1} but for $\phi$=135\degree. 
    Left: Stokes vector at 1~km altitude for downward viewing directions. 
    Right: Absolute difference between individual
    models and MYSTIC, the grey area corresponds to 2$\sigma$.
    Arrows indicate out-of-range values.} 
  \label{fig:b4_results2}
\end{figure}
\begin{figure}[t]
  \centering
  \includegraphics[width=1.\hsize]{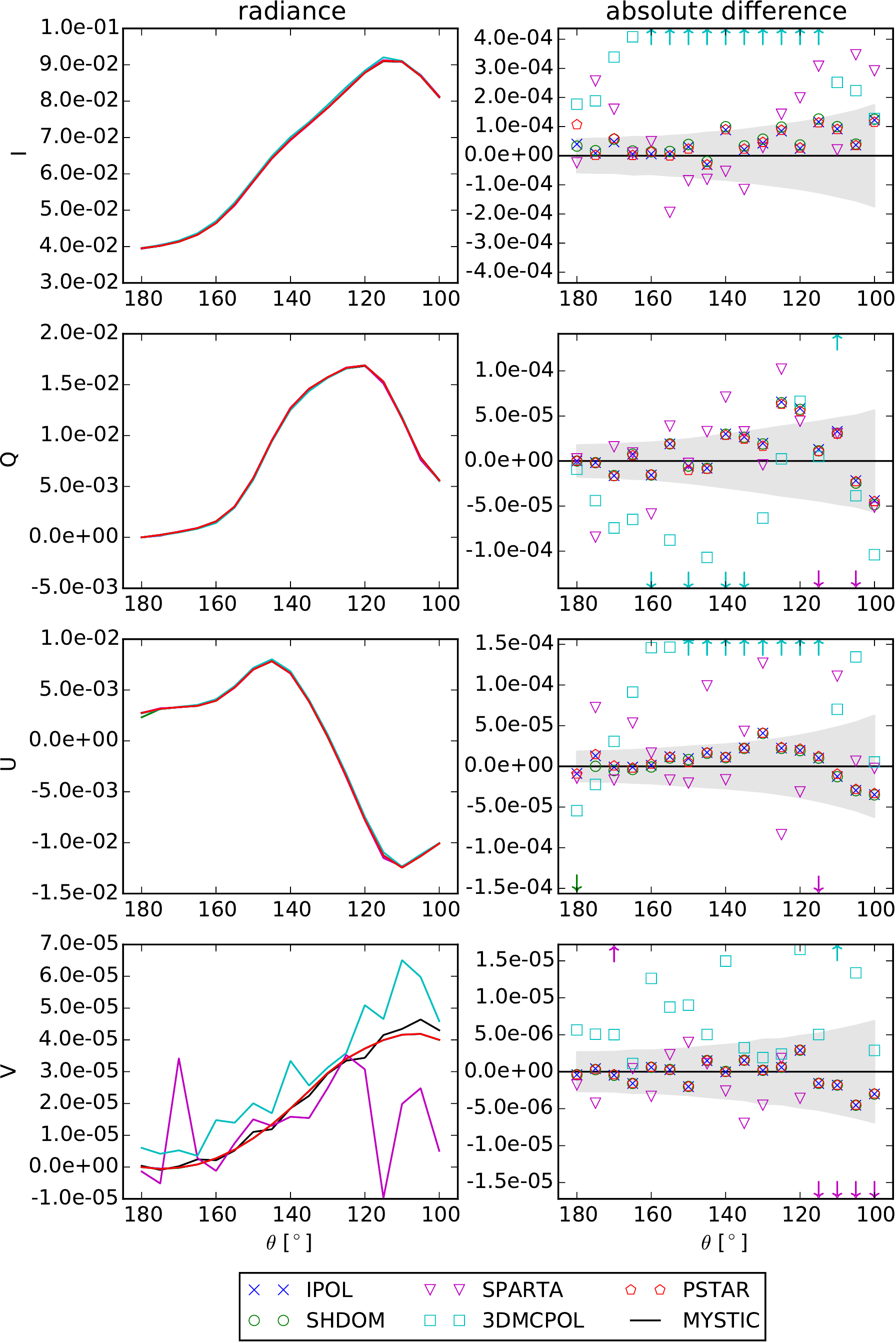}
  \caption{Test case B4, settings as Fig.~\ref{fig:b4_results2}. 
    Left: Stokes vector at the top of the atmosphere. 
    Right: Absolute difference between individual
    models and MYSTIC, the grey area corresponds to 2$\sigma$.
    Arrows indicate out-of-range values.} 
  \label{fig:b4_results3}
\end{figure}
For the most demanding case of a cloud layer above an ocean surface
embedded in a Rayleigh atmosphere we find the largest
differences between the models. 
Figs.~\ref{fig:b4_results1}--\ref{fig:b4_results3} show the radiance
field and the differences between the models. 

Fig.~\ref{fig:b4_results1} shows the radiance field for a viewing azimuth 
angle of 0\degree\ at the surface. Here all models
agree quite well when the forward scattering direction is excluded,
only the 3DMCPOL result for $Q$ is different from other models.
This difference is mainly due to the error in the implementation of
surface reflection in 3DMCPOL.
Within the 2$\sigma$ range the models Pstar, IPOL and SHDOM mostly
agree to MYSTIC. There is one outlier in the SHDOM results for $Q$ at
a viewing zenith angle of 0\degree. In the exact forward direction Pstar is
more than a factor of 2 larger than the other models. This is because
Pstar uses a relatively small number of streams, i.e. 60 for this
case. The result in exact forward direction improves by increasing the number of
streams. 
The value of the forward scattering peak agrees for MYSTIC, IPOL and
SPARTA.  

Fig.~\ref{fig:b4_results2} shows the radiance field at a viewing azimuth 
angle of 135\degree\ for down-looking directions at 1~km altitude, below the cloud layer,
We find that the models MYSTIC, SHDOM and Pstar agree for $I$, $Q$ and $V$.
For $U$ SHDOM and MYSTIC are negative whereas Pstar is
slightly positive. 

Fig.~\ref{fig:b4_results3} shows the radiance field at a viewing azimuth 
angle of 135\degree\ for down-looking directions at the top of the
atmosphere, where we see part of the cloudbow. 
IPOL, SHDOM and Pstar mostly agree to MYSTIC within the 2$\sigma$
range. SHDOM again shows an outlier at 180\degree viewing zenith angle. 
SPARTA results are more noisy than MYSTIC but there are no
obvious systematic differences and within the
uncertainty the models agree. For 3DMCPOL the differences are a bit
larger and systematic, e.g. for $I$ the 3DMCPOL results are
systematically larger than other model results.

Tab.~\ref{tab:b_results} shows that the relative root mean square
difference between MYSTIC and all other models
is very small for $I$, the largest difference is 0.7\%.
For $Q$ and $U$,  $\Delta_m$ is
smaller than 0.6\% for IPOL,  1--4\% for SHDOM, Pstar and SPARTA,  
and 5--27\% for 3DMCPOL. For $V$, $\Delta_m$ is about 20\%
for IPOL, SHDOM and Pstar; this large value is due to the noisy MYSTIC
result. Since 3DMCPOL and SPARTA are more noisy than MYSTIC the values
are even larger. 

\section{Conclusion and Outlook}

Overall, we found a very good agreement between all models and for
the test cases of this intercomparison project. 
The achieved level of agreement is very high, for cases without
clouds the relative root mean square difference is mostly below 0.05\%
for total intensity and linear polarization. 
However some
significant deviations were found: for non-zero depolarization
factors, for ocean reflection, and for simulations including cloud
droplets. For these
settings some of the models need to be corrected or improved. 

For all single layer calculations we found an agreement between
the models MYSTIC, IPOL, SPARTA and Pstar.
SHDOM also agrees for all cases and almost all viewing directions, but
for the cloud layer cases it shows artefacts at a few specific viewing
directions.
3DMCPOL agrees well for Rayleigh scattering with depolarization factor
set to 0, for the Lambertian surface, and for aerosol and cloud
cases. There are small differences when the depolarization
factor is non-zero and also for the case with ocean reflectance
matrix, for these two cases we may conclude that 3DMCPOL is not
consistent with other models and should be corrected. 

For the multi-layer cases the radiance fields at the surface and at
the top of the atmosphere were provided by all participants. 
We again find a perfect agreement between MYSTIC, IPOL, SPARTA and
with a few exceptions SHDOM, which shows artefacts at a few specific
angles for the cloud case,
in particular at viewing angles of 0\degree and 180\degree.
Pstar agrees for all cases except the last one with ocean reflectance
matrix and cloud layer, where we find small deviations at specific
geometries for the $U$
component of the Stokes vector.
3DMCPOL shows small differences for all cases due to the
non-zero depolarization factor of 0.03 which was used for all
multi-layer cases. Larger differences appear when absorption is taken
into account, here the 3DMCPOL model should be improved. 
Also for the cloud layer above ocean surface we find larger
differences, as expected because we have already seen these differences
in the single layer case with ocean reflectance matrix. 
 
As benchmark we provide the MYSTIC results, which agree to IPOL and
SPARTA for the delivered cases at the surface and the top of the
atmosphere. MYSTIC also agrees to SHDOM for most viewing directions,
the exceptions are obviously due to artefacts in SHDOM at specific
angles. Also it agrees 
to Pstar for cases without ocean reflectance matrix.
Along with the radiance data we
provide the standard deviations which are helpful when developers want to
use the benchmark data for testing their models.

The detailed setup for all cases, the benchmark results as well as
plots showing model results of all cases are publically available at
the IPRT website
(\url{http://www.meteo.physik.uni-muenchen.de/~iprt}).

The next phase of the intercomparison project will start in spring
2015. We will then focus on three-dimensional scenarios including
clouds and aerosols. 

\section*{Acknowledgement}
We thank Dr. Michael Mishchenko for providing the code to calculate the ocean reflectance matrix. 
Furthermore we thank Dr. Josef Gasteiger for providing optical properties of
the aspherical aerosol particles.

\bibliographystyle{elsarticle-num-names}
\bibliography{./literature.bib}

\end{document}